\RequirePackage{fix-cm}
\documentclass[twocolumn]{svjour3}          %
\smartqed  %

\usepackage{graphicx}
\usepackage{xspace}
\usepackage{xurl}
\usepackage{listings}
\usepackage{microtype}
\usepackage[hidelinks]{hyperref}
\usepackage{pgfplots}
\usepackage{upquote}           
\usepackage{inconsolata}

\usepgfplotslibrary{groupplots}
\usetikzlibrary{calc,arrows.meta}
\pgfplotsset{compat=1.18}

\definecolor{cProg}{HTML}{1B6CA8}
\definecolor{cMeta}{HTML}{D1793B}
\definecolor{cExec}{HTML}{B23A48}
\definecolor{cRead}{HTML}{2E7D5B}
\definecolor{cNeutral}{HTML}{8C8C8C}

\pgfplotsset{
  delilah base/.style={
    width=\linewidth,
    grid=major,
    major grid style={line width=0.2pt, draw=gray!30},
    tick align=outside,
    tick pos=left,
    label style={font=\footnotesize},
    tick label style={font=\footnotesize},
    legend style={font=\footnotesize, draw=none, fill=none},
    legend cell align={left},
  },
  delilah log ms/.style={
    log ticks with fixed point,
    /pgf/number format/1000 sep={},
    ylabel={Latency (ms, log scale)},
  },
  delilah line/.style={line width=0.9pt, mark size=1.8pt},
}

\lstdefinestyle{base}{
  basicstyle=\ttfamily\footnotesize,
  columns=fullflexible,         %
  keepspaces=true,
  upquote=true,
  numbers=left, numberstyle=\tiny\color{gray}, numbersep=6pt,
  xleftmargin=1.5em,
  frame=tb, framerule=0.4pt,
  showstringspaces=false,
  tabsize=2,
  breaklines=true,
  keywordstyle=\bfseries,
  commentstyle=\itshape\color{gray},
  stringstyle=\color{teal},
  captionpos=b,
}

\lstdefinelanguage{ebpfc}{
  language=C,
  morekeywords={uint8_t,uint16_t,uint32_t,uint64_t,int32_t,int64_t,
                u8,u16,u32,u64,s32,s64,size_t,bool,
                __u8,__u16,__u32,__u64,inline,__always_inline},
  emph={filter_eq_neon,filter_bw_neon,regfunc},
  emphstyle={\color{blue!60!black}},
  emph={[2]vdupq_n_u32,vld1q_u32,vst1q_u32,vcgeq_u32,vcleq_u32,
         vandq_u32,vorrq_u32,vshlq_n_u32,uint32x4_t},
  emphstyle={[2]\color{violet}},
  morecomment=[l][\color{gray}]{\#pragma},
}
\lstdefinestyle{ebpfc}{style=base, language=ebpfc}

\lstdefinelanguage{ebpfasm}{
  morekeywords={call,exit,goto,if,ll,be16,be32,be64,le16,le32,le64},
  morekeywords=[2]{u8,u16,u32,u64,s8,s16,s32,s64},
  keywordstyle=[2]\color{teal},
  emph={r0,r1,r2,r3,r4,r5,r6,r7,r8,r9,r10,w0,w1,w2,w3,w4,w5,w6,w7,w8,w9},
  emphstyle={\color{blue!60!black}},
  sensitive=true,
  morecomment=[l]{\#},
  morecomment=[l]{;},
  morecomment=[l]{//},
  alsoletter={.},               %
}
\lstdefinestyle{ebpfasm}{style=base, language=ebpfasm, numbers=left}

\lstdefinelanguage{nativec}{
  language=C,
  morekeywords={uint8_t,uint16_t,uint32_t,uint64_t,int32_t,int64_t,
                size_t,ssize_t,bool,inline,static,restrict,const},
  emph={uint32x4_t,uint8x16_t,uint16x8_t,uint64x2_t,int32x4_t,
        vdupq_n_u32,vld1q_u32,vst1q_u32,vcgeq_u32,vcleq_u32,vceqq_u32,
        vandq_u32,vorrq_u32,vshlq_n_u32,vaddq_u32,vgetq_lane_u32},
  emphstyle={\color{violet}},
  morecomment=[l][\color{gray}]{\#},
}
\lstdefinestyle{nativec}{
  style=base,
  language=nativec,
  backgroundcolor=\color{gray!8},   %
}

\newcommand{\delilah}{\texttt{Delilah}\xspace}

\makeatletter\def\makeheadbox{}\makeatother

\begin{document}

\title{Query Processing with eBPF-offload on Computational Storage}

\author{Alexander Krause \and
        Niclas Hedam \and
        Johannes Pietrzyk \and
        Dirk Habich \and
        Wolfgang Lehner \and
		Sangjin Lee \and
		Alberto Lerner \and
		Philippe Bonnet }

\institute{A. Krause, J. Pietrzyk, D. Habich, W. Lehner \at TU Dresden \\
           \email{\{alexander.krause, johannes.pietrzyk, dirk.habich, wolfgang.lehner\}@tu-dresden.de}
           \and
           N. Hedam \at IT University of Copenhagen \\ \email{orcid@sl.hed.am}
           \and
           S. Lee, A. Lerner \at Computing Flows \\ \email{\{sangjin, lernera\}@computingflows.com}
           \and
           P. Bonnet \at University of Copenhagen \\ \email{bonnet@di.ku.dk} \\ ORCID: 0000-0002-0234-5374 
}
\date{}

\maketitle

\begin{abstract}
The idea of pushing computation to storage devices has been explored for decades.
In recent years, research has focused on computational storage, i.e., devices that allow a host
to execute functions as I/O commands on local data.
Today, the NVMe and SNIA computational storage standards share an execution model in which the host invokes
functions on the device, on data that the host has moved into device memory. Both standards single out eBPF
as a format for functions offloaded by the host. 
In this paper, we study whether this form of computational storage is a good match for query processing. 
We present the design and implementation of \delilah, a prototype based on an open-hardware device that implements
the standard execution model 
on an actual device and runs eBPF programs calling device-resident functions.
As an initial step, we hand-coded a portion of SSB Q3.3 for execution on the device,
aiming to integrate computational storage with a database system's storage manager. 
The series of experiments that we ran showed that 
offloading gains little over host execution, execution degenerates into column-at-a-time processing,
and time is spent staging and moving data in device memory.
Because these roadblocks stem from the execution model rather than from the query, we did not pursue integration further.
We detail these limitations and lay the groundwork for a different computational storage model better aligned with data-intensive workloads.

\end{abstract}

\section{Introduction}
\label{sec:introduction}

To avoid the data-access bottleneck~\cite{mutlu_intelligent_2021}, data-centric architectures
place processing capability near where the data resides. 
For data management, this idea goes back to active disks~\cite{riedel_active_2001} and 
has resurfaced with each generation of storage hardware~\cite{balasubramonian_near-data_2014}.
Its current incarnation is computational storage, a form of near-data processing that enables hosts
to invoke functions on device data via I/O commands~\cite{Lerner2024-kb}.
After a first generation of prototypes and products based on proprietary interfaces, computational storage is now standardized 
through 
SNIA\footnote{Storage Networking Industry Association, the trade association representing storage companies.}~\cite{Storage_Networking_Industry_Association2026-zn,Storage_Networking_Industry_Association2025-ox}
and NVMe\footnote{Non-Volatile Memory Express, the consortium defining the industry-standard for Solid-State Drives.}~\cite{NVM-Express2025-uz,NVM_Express2025-mb}.

The SNIA programming model is broad~\cite{Storage_Networking_Industry_Association2026-zn}. 
In its {\em direct} usage model, the host invokes a function on data in device memory. 
In its {\em indirect} usage model, the storage controller moves data from storage into device memory and triggers 
a function as a side effect of a storage request. 
{\em Aggregator} functions invoke sequences of functions on the device, possibly in parallel and possibly 
repeatedly, so as to operate on data with minimal host involvement.

The SNIA and NVMe interfaces that realize this model are much narrower~\cite{Storage_Networking_Industry_Association2025-ox,NVM-Express2025-uz}.  
They are both explicitly tailored for a host-orchestrated interface. 
In NVMe, a computational program operates only on data in the device memory, denoted subsystem local memory~\cite{NVM_Express2025-mb}. 
The host copies data into that memory and executes the program with separate commands. 
The SNIA API~\cite{Storage_Networking_Industry_Association2025-ox} softens this model with batch requests, which chain storage, copy and compute requests into 
an acyclic graph that the device executes without interacting with the application. 
A batch is fixed when it is submitted, so it cannot contain loops, and its structure cannot depend on the data it processes. 
In short, computational storage, as exposed through standard interfaces, is based on 
{\em host-invoked functions} and {\em host-managed data movement}. 
The device performs the computation; the host orchestrates it.

This model is familiar.
It is an accelerator-offload model in which a host launches kernels over buffers it explicitly populates.
It is also attractive from a standardization standpoint: it adds a few operations to the storage interface while leaving the storage stack semantics untouched.
The question we address in this paper is whether it is a good match for query processing. 

There are reasons to doubt. 
First, computational storage relies on computation that accesses SSDs through their external interface~\cite{Lerner2024-kb}. 
As a result, computational storage has no bandwidth or latency advantage over the host CPU when accessing stored data~\cite{lerner_not_2021}.
Since computational storage increases the cost of the storage subsystem, it must pay off in other ways: 
by reducing the amount of data transferred to the host, by freeing host cores and memory, 
or by executing hardware-accelerated functions~\cite{do_better_2021,kwon_hardwaresoftware_2022}. 

Second, device memory is small and shared by data staged from the SSDs, 
intermediate results and results to be transferred to the host: memory, not compute, 
is the scarce resource on the device.
Query processing stresses exactly these constraints. 
We cannot assume that base tables fit in device memory, so columns must be processed in chunks. 
Operators maintain state across chunks (e.g., selection vectors), 
and the amount of data flowing between them depends on selectivities that are only known at run-time. 
Previous work characterized offloadable functions as small, I/O-bound compute kernels that account for a 
significant portion of an application's execution time~\cite{wilcox_design_2021}. 
Whether query operators, orchestrated from the host, fit this characterization is an open question.

Both SNIA and NVMe single out extended Berkeley Packet Filter (eBPF) as a favoured
format for representing functions downloaded at run-time: 
SNIA lists eBPF among its execution environments, and its API defines eBPF programs as a type of downloadable 
function~\cite{Storage_Networking_Industry_Association2025-ox}; 
NVMe proposes eBPF as the representation of downloadable programs~\cite{NVM-Express2025-uz}. 
eBPF originates in the BSD Packet Filter, designed to evaluate boolean-valued functions 
without copying packets inside a UNIX kernel~\cite{mccanne_bsd_1993}. 
Modern eBPF~\cite{starovoitov_lkml_nodate} executes user-defined functions inside the Linux kernel for 
performance monitoring~\cite{gregg_bpf_2019}, security monitoring~\cite{fournier_runtime_2021} 
and storage management~\cite{zhong_xrp_2022}, and serves as an intermediate representation for 
hardware-accelerated packet processing~\cite{brunella_hxdp_2022}.
It would be tempting to judge eBPF for computational storage by the standards of packet filtering, 
as a sandboxed language in which stateless functions of a few instructions are executed on data streaming 
through a device, or to compare it with other bytecodes as a general-purpose data processing 
language~\cite{huang_evaluation_nodate}. 
We argue that this is not its role in computational storage.

In computational storage, eBPF is best understood as a {\em relocatable binary format}. 
eBPF is a vendor-neutral instruction set~\cite{hedam_delilah_2023}, generated from C by both gcc and clang. 
An eBPF program refers to external functions by name; the program resolves these references when it is linked against
the functions registered with the run-time environment on the device.
Device-defined functions, possibly hardware accelerated, can thus be registered with the eBPF run-time 
environment and called from a program generated on the host, without extending the host-device interface 
with a new operation for each function (Section~\ref{sec:designspace}).
In this role, eBPF programs act as glue: they massage the inputs and outputs of registered functions and orchestrate their execution.
At the same time, native or hardware-accelerated code performs most of the processing. eBPF's limitations as a programming language
(no floating point, no indirect calls, and the difficulty of verifying programs with loops) matter less in this role. However, safely
executing downloaded code remains an open problem.

In SNIA terms, an eBPF program that calls registered functions is an aggregator function: 
it runs in an eBPF environment and invokes functions that may run in other environments, e.g., in hardware. 
Unlike a batch request, it can contain loops and make data-dependent decisions. 
eBPF does not change the execution model. Through standard interfaces, an eBPF program is a 
host-invoked function, executed on data the host has moved into device memory, 
and the sequence it orchestrates is bounded by what the host has staged. 
eBPF determines {\em what} to offload and how to package it; 
the execution model determines {\em how} offloaded functions interact with data. 
It is the latter that we put to the test.

To this end, we design and implement \delilah, a computational storage system that follows the host-orchestrated execution model
of standard interfaces on an actual open hardware device, the OpenSSD Daisy platform~\cite{daisy-sk}, using an on-path PCIe architecture. 
\delilah runs eBPF programs in user space on the device's ARM cores, links them against registered functions, and exchanges programs,
arguments, and results with the host through program and data slots in device memory.
We use \delilah\ to evaluate the mechanisms on
which the execution model rests: the cost of invoking an offloaded function, the coherence of the buffers shared between host and device,
and the transfer of data between them.
We then map query processing onto this execution model: we split columns into chunks that fit in device memory, register operators as
functions invoked chunk by chunk from the host, and keep two chunks in flight to overlap processing and transfers.

As a first step, we hand-code query Q3.3 from the Star Schema Benchmark (SSB). This is a sanity check before integrating \delilah\ into a database system's
storage manager. We chose Q3.3 because it is a selective query over a few columns with simple operators,
so it should be well suited for offloading on computational storage.
On this query, we run a campaign of experiments that vary what is offloaded, how filters are
implemented (Neon SIMD or FPGA),
and how operators are scheduled, and we compare against host-only execution.

This first step was also our last. We identified showstoppers that no amount of query tuning can work around,
because they stem from the execution model itself: functions are invoked by the host, and data staging and
data movement are both managed by the host. These showstoppers already appear in the most favorable setting
we could construct, on a data set whose columns fit entirely in device memory. Larger data sets, more columns, and more complex plans would only multiply the per-chunk round trips and
staging passes that already dominate execution time. We therefore report these negative findings rather than
pursue the integration further.

Query processing calls for an execution model in which the device manages the iteration over data and 
the movement of data between SSD and device memory. 
The SNIA programming model anticipates device-side orchestration with indirect usage and variable-sequence aggregator functions, 
but the standard interfaces do not yet expose them. 
However, we argue that device-side orchestration is not enough.
We propose a more holistic approach to SSD programming that addresses computational storage's shortcomings.

In summary, our contributions are the following:
\begin{itemize}
	\item We survey computational storage standards and clarify eBPF's role in computational storage. 
	\item We describe the design and implementation of \delilah, a computational storage system following the 
	host-orchestrated execution model on a real device (Section~\ref{sec:delilah}), 
	and a chunked, column-at-a-time execution model for offloaded query processing (Section~\ref{sec:qp}).
	\item We evaluate eBPF offload overhead and host-device data transfers on \delilah.
	Through a campaign of experiments on a hand-coded SSB query, we analyze where time goes when
	offloading (Section~\ref{sec:evaluation}). We show that the limitations we encounter are structural, and conclude that
	host-invoked functions and host-managed data movement are not a good match for query processing.
	\item We draw the lessons from this work and discuss perspectives for near-data processing with co-designed SSDs.
\end{itemize}

This article extends our DaMoN'23 paper~\cite{hedam_delilah_2023}, which introduced \delilah. 
The new material provides a detailed analysis of the revised SNIA and NVMe standards,
details of the \delilah\ system
including the controller, driver and block design, 
and the offload of query processing, including the chunked execution model and its evaluation with an SSB query.
Code and experiments are available online
at \url{https://github.com/delilah-csp}.

\section{Related Work}
\label{sec:related}

Computational storage~\cite{barbalace_computational_2021} is the current incarnation of decades-old ideas 
about active storage~\cite{riedel_active_2001} and near-data processing~\cite{balasubramonian_near-data_2014}.
It refers to architectures where storage functions are executed, as a result of I/Os, 
on compute resources located outside Solid-State Drives. 
Computational storage functions thus only have access to the external interface of SSDs. 
This distinguishes computational storage from programmable and co-designed SSDs, 
where programs have access to internal SSD interfaces~\cite{Lerner2024-kb}.

We organize related work along two dimensions: execution model and code representation. 
The {\em execution model} determines who invokes offloaded functions, and who moves the data they operate on. 
{\em Code representation} concerns how offloaded functions are represented. 
Here, we focus on eBPF.

\subsection{Execution Model}
\label{sec:related:execution}

\paragraph{Proprietary interfaces.}
The first generation of commercial computational storage used proprietary interfaces, such as 
ScaleFlux~\cite{cao_polardb_2020-2}, NGD~\cite{heydarigorji_leveraging_2022},
Samsung SmartSSD~\cite{samynathan_computational_2019}, or Seagate Kinetic Drive~\cite{montana_moveable_2022}.
In particular, Cao et al.~\cite{cao_polardb_2020-2} tackle analytical workloads in the cloud-native 
relational database POLARDB with ScaleFlux computational storage.
Each device defined its own interface, its own set of offloadable functions, and its own division of labor 
between host and device---the SNIA and NVME standardization efforts aimed at reducing this diversity.

SNIA defines a programming model for computational storage, together with an application programming interface; 
NVMe defines the device interface. 
We review them in turn.

\paragraph{SNIA programming model.}
The SNIA Computational Storage Architecture and Programming Model~\cite{Storage_Networking_Industry_Association2026-zn}, 
first released in 2022 and approved as version~1.2 in April 2026, defines the terminology, architecture and programming 
model of computational storage. It makes no assumption about the physical interface between host and device. 
It distinguishes three kinds of computational storage devices (CSx): 
computational storage processors (CSP), which provide computation for an associated storage system without persistent storage; 
computational storage drives (CSD), which also contain persistent storage; 
and computational storage arrays (CSA). 
The computational storage resources of a CSx comprise:
\begin{itemize}
    \item computational storage engines (CSE), i.e., components such as a CPU or an FPGA that execute computational storage functions (CSF);
    \item computational storage engine environments (CSEE), i.e., the operating environment of a CSE, 
    e.g., an operating system, a container platform, an eBPF environment or an FPGA bitstream;
    \item a resource repository that holds CSFs and CSEEs available for activation;
    \item function data memory (FDM), portions of which are allocated (AFDM) to individual instances of CSF operations. 
    An instance may only access its own device memory.
\end{itemize}
CSEEs and CSFs are either pre-installed or downloaded by the host into the resource repository. 

The programming model defines two usage models. 
In the {\em direct} usage model, the host sends a request that invokes a CSF. 
In the reference flow, the host places the input data in AFDM before the request, and the CSE places its results in device memory; 
the model also allows inputs and outputs to reside in device storage or in host memory. 

In the {\em indirect} usage model, the host configures the storage controller to associate a CSF with storage requests, 
based on request parameters, on an address range or on data characteristics. 
The storage controller then moves data from storage into device memory, triggers the CSF, and returns its results in response to the storage request. 

Version~1.1 added {\em sequencing}, to operate on data with minimal host involvement. 
An aggregator CSF invokes a sequence of activated CSFs on the same device, sequentially or in parallel, 
possibly across CSEEs. 
A fixed-sequence aggregator hard-codes the sequence. A variable-sequence aggregator determines it from parameters and 
intermediate results, possibly invoking the same CSF repeatedly.

One example of the standard splits a large input into chunks that parallel search CSFs process before combining their results. 
The standard also lists example functions that are directly relevant to query processing, 
including a database filter CSF (projection and selection) and a pipeline CSF that combines commands according to a data flow specification.
The SNIA programming model is thus broader than host-invoked functions and host-managed data movement: 
indirect usage delegates data movement to the storage controller, 
and aggregator CSFs move the orchestration of functions onto the device.

\paragraph{SNIA API.}
The SNIA Computational Storage API~\cite{Storage_Networking_Industry_Association2025-ox}, a SNIA standard whose version~1.1 was released in May 2025, 
defines the interface between applications and computational storage devices. 
It is agnostic to how devices are attached (directly, over a network or over a fabric): 
a library, extended with plugins, maps API functions onto the specific interface of each device, e.g., NVMe. 
A plugin may also substitute host CPU execution for a device that is absent. 
Besides discovery and configuration functions, which activate CSEEs and CSFs, the API defines functions 
to download CSFs (including eBPF programs, one of the CSF types of the API) and four groups of functions 
that structure the use of a device:
\begin{itemize}
    \item {\em memory management} functions to allocate and free device memory;
    \item {\em storage requests} load data from storage into device memory, or store data from device memory to storage, 
    at the block or file level, possibly with peer-to-peer transfers that bypass host memory;
    \item {\em copy requests} move data between host memory and device memory, or within the device;
    \item {\em compute requests} execute an activated CSF, whose arguments are scalars or references to device memory.
\end{itemize}
All requests are queued and complete synchronously or asynchronously (via a callback or an event).
The reference flow follows the direct usage model of the programming model: the application allocates device memory 
for inputs and outputs, queues a storage request to load data, queues a compute request, and queues a copy request 
to retrieve the results.
The API is explicit about the resulting division of labor. It is tailored for a host-orchestrated interface.
Even when data is transferred 
peer-to-peer between SSD and AFDM, the host remains in charge of orchestrating it. 
Correspondingly, the API defines no function that associates a CSF with storage requests, as the indirect
usage model would require.

The API mitigates the cost of host orchestration with {\em batch requests}. 
A batch combines storage, copy and compute requests, executed serially, in parallel, 
or as a graph of dependencies (hybrid mode), and the device executes it without interacting with the application. 
Batches may be supported in hardware or emulated in software. 
A batch is, however, fixed when it is submitted: an entry that would create a loop is rejected, 
the number of entries is bounded by the device, and the structure of the batch cannot depend on the data it processes. 
Batch requests thus implement a form of fixed-sequence aggregation, defined by the host, 
but not the variable-sequence aggregation of the programming model.

\paragraph{NVMe.}
\label{sec:nvme}
NVMe defines the device interface for computational storage 
with two command sets, Computational Programs (TP~4091) 
and Subsystem Local Memory (TP~4131), now ratified~\cite{NVM-Express2025-uz,NVM_Express2025-mb}. 
They correspond to CSFs and FDM in the SNIA model. 
Computational programs operate only on data in subsystem local memory (SLM). 
The host loads or activates a program on a computational programs namespace, 
describes the ranges of SLM a program operates on as a memory range set, 
and executes the program with a command that references this memory range set. 
Data is moved between NVM namespaces and SLM with copy commands, 
and between SLM and the host with memory read and write commands, all issued by the host. 
The structure is the same as with the SNIA API: data is read from an SSD into SLM by the host prior to program execution, 
and results are read (or written to an SSD) by the host after program execution. 
NVM Express itself describes its Computational Programs command set as a host-driven approach. 
NVMe distinguishes between two kinds of programs. 
Device-defined programs are installed in hardware (FPGA bitstream) or software (OS image) before a namespace is attached. 
Downloadable programs, represented in a hardware-agnostic bytecode such as eBPF, are loaded on a 
computational programs namespace by the host at run-time.

\paragraph{Host-invoked functions and host-managed data movement.}
The two standard interfaces thus agree, at the application level and at the device level: 
functions are invoked by the host on data that the host has moved into device memory, 
and results are moved by the host out of device memory. 
This is the execution model that we refer to as {\em host-invoked functions and host-managed data movement}. 
It is the model that standard computational storage interfaces expose today, and the one we evaluate in this paper. 
Designing and evaluating a device that supports the elements of the SNIA programming model that go beyond it,
indirect usage (which can be seen as a form of database kernel~\cite{dbkernel24-de}) and variable-sequence aggregation, 
remains a topic for future work.

\paragraph{Prototypes.}
Samsung demonstrated a second generation SmartSSD~\cite{noauthor_samsung_nodate} supporting eBPF offload over NVMe at SC'22.
Among systems described publicly, ZCSD~\cite{lukken_zcsd_2021} offloads eBPF bytecode to a computational storage 
device over ZNS SSDs emulated in QEMU, via a local function call. In a real system, this mechanism must be replaced 
by a protocol between host and device over PCIe or fabric.
The Eid-Hermes project~\cite{oliveira_eid-hermes_2022}, led by Eideticom, defines such a protocol for eBPF offload over PCIe. 
It enables a host to load an eBPF program to a PCIe device, run the program synchronously and then unload it.
The eBPF program is executed in user-space with the uBPF virtual machine\footnote{\label{f:ubpf}\url{https://github.com/iovisor/ubpf}}.
Parameters are written by the host to a buffer on the device, and results are explicitly read back by the host: 
Eid-Hermes is a direct instance of host-invoked functions and host-managed data movement.
It is implemented in QEMU. We use it as a point of departure for \delilah, 
which implements this execution model on an actual computational storage device (Section~\ref{sec:delilah}). 
The host interface of \delilah\ covers, in a simplified form, the memory, copy and compute requests of the SNIA API.
The key difference is that \delilah\ supports registered functions that read data from the underlying SSD
into the device memory.

Wilcox et al.~\cite{wilcox_design_2021} consider an architecture for code offload on computational storage, 
also emulated in QEMU, in which compute kernels are loaded and then executed asynchronously. 
Compute kernels can call a restricted library of pre-defined SSD access functions, i.e., they can initiate I/O 
from the device, which departs from purely host-managed data movement.
Wilcox et al. also characterize the code worth offloading: 
{\em a compute kernel is a small section of code which is both I/O bound and comprises a significant portion 
of the overall execution time of the application}. A compute kernel must be small enough to be efficiently 
executed on the device, I/O bound to benefit from offloading to storage, and significant enough for its offload 
to have an impact on performance. 
Our evaluation of query processing (Section~\ref{sec:eval:qp}) shows that this characterization 
is not sufficient: individual operators may well be small, I/O bound and significant, yet the way they are 
orchestrated from the host determines where time goes.

\paragraph{Query processing near storage.}
Offloading query processing to storage has been studied extensively on programmable SSDs, 
where code runs on the SSD controller and reads flash through internal interfaces, 
i.e., on co-designed SSDs in the terminology of Lerner et al.~\cite{Lerner2024-kb}.
Do et al.~\cite{do_query_2013} push selection and aggregation from Microsoft SQL Server into a Samsung Smart SSD. 
The operators are compiled into the SSD firmware and exposed as commands; the host opens a session, 
the device processes the data as it is read from flash, and the host retrieves the results. 
Joins are not offloaded, and the data layout had to be converted to obtain speedups.
Biscuit~\cite{gu_biscuit_2016} generalizes this approach into a framework for a commodity NVMe SSD. 
Its programming model is based on data flow: tasks that run on the SSD (SSDlets, written in C++ and loaded at run-time) 
or on the host communicate through typed ports. The host loads a module, instantiates SSDlets, connects their ports 
into a graph and starts the application; SSDlets then read the files passed to them by the host 
through the internal I/O path, and data streams through the ports. 
Communication between host and device is expensive (hundreds of microseconds per port transfer), 
so the framework favors long-running streaming tasks over fine-grained invocations.
YourSQL~\cite{Jo2016-nq} ports MariaDB onto Biscuit and offloads early filtering. 
The query planner selects a table with restrictive predicates, estimates its page-level selectivity with a sampling task 
on the SSD, and places the table first in the join order. 
Filter tasks on the SSD then scan the entire table without interruption, using hardware pattern matchers 
on the flash channels and a software filter, and accumulate per-page match hints. 
The host pulls these hints, reads the qualifying pages in bulk and executes the rest of the query, 
in a pipeline that overlaps filtering on the device with processing on the host. 
In a micro-benchmark on a column-store layout of TPC-H tables, YourSQL also observed that page-level early 
filtering is ineffective on an uncompressed column store, where almost every page contains a qualifying value.

In all three systems, the device, not the host, drives the iteration over stored data: 
the host intervenes once per scan or per query, not once per chunk. 
Biscuit, in particular, demonstrates on a commercial NVMe SSD the kind of execution model that 
the SNIA programming model anticipates with variable-sequence aggregation and device-managed data movement, 
i.e., a host-defined graph of device-side tasks that move data themselves. It does so through a proprietary interface. 
Two differences with computational storage must be kept in mind. 
First, these systems benefit from the internal bandwidth of the SSD, which exceeds the bandwidth of its host interface: 
Do et al. relied on a gap of almost 3$\times$ with SATA/SAS SSDs, and Biscuit reports roughly 1~GiB/s of additional 
internal bandwidth over PCIe Gen3$\times$4. 
A computational storage device accesses SSDs through their external interface and has no such advantage.
Second, they are tied to a specific SSD and its firmware. 
Our question is thus different: Can query processing benefit from computational storage as exposed by the 
standard interfaces, with host-invoked functions and host-managed data movement over the external SSD interface.

\paragraph{Data-driven offload.}
Outside computational storage standards, several lines of work offload code with an execution model 
in which the arrival of data or requests, rather than the host, drives execution. 
XRP~\cite{zhong_xrp_2022} executes eBPF storage functions in the NVMe driver of the Linux kernel. 
A function can resubmit I/Os from the driver, e.g., to traverse an on-disk index, 
which avoids a round trip through the storage stack for each step of the traversal.
Code offload on smart NICs has also matured rapidly. 
Smart NICs, i.e., network interfaces that allow stateful computation on messages or flows, play a central role in 
the convergence of hyperscale data center and high-performance computing networks~\cite{hoefler_convergence_2022}. 
A prime example is sPIN~\cite{hoefler_spin_2017}, an architecture supporting the execution of
user-defined functions within a smart NIC. sPIN distinguishes between header, payload and completion handlers 
that are invoked as the corresponding parts of a message stream through the NIC. 
sPIN was recently used for offloading storage policies, including erasure coding, 
for disaggregated storage over RDMA~\cite{di_girolamo_building_2022}. 
In this context, throughput is non-negotiable (e.g., 200 or 400~Gb/sec): 
the execution of offloaded code cannot interfere with the expected NIC throughput, 
which restricts user-defined functions to a few instructions per packet, typically focused on
data movement~\cite{Ryser2022-wo} .

\paragraph{DPUs and SSDs.}
Data Processing Units (DPUs), i.e., SoC-based smart NICs with general-purpose cores, memory and accelerators, 
are increasingly attached directly to SSDs.
In {\em SmartNIC JBOFs}, a smart NIC replaces the server in front of a set of NVMe SSDs. 
Gimbal~\cite{Min2021-xh} implements a software storage switch on such a smart NIC that schedules I/O between 
network ports and SSDs for co-located tenants. 
LEED~\cite{Guo2023-ev} runs a persistent key-value store on an array of SmartNIC JBOFs, 
and Scalio~\cite{Sun2025-fz} scales a DPU-based JBOF key-value store by serving reads through RDMA and 
NVMe-oF target offload rather than through the DPU cores. 
In database systems, DDS~\cite{Zhang2024-jd} uses a DPU in front of the SSDs of a disaggregated storage server: 
client read requests are executed directly on the DPU, which accesses the SSDs without involving the host, 
and an offload engine executes user-defined requests on the DPU, while updates remain on the host because DPU memory is small. 
DPDPU~\cite{Hu2024-jo} generalizes this approach into a platform with compute, network and storage engines 
that expose asynchronous interfaces to data systems. 
Thostrup et al.~\cite{Thostrup2022-lq} characterize BlueField DPUs from the perspective of a DBMS on fast networks.

These systems share two traits with our work. 
First, execution on the DPU is driven by requests arriving from the network, 
and the DPU, not the host, moves data between SSDs, its own memory and the network. 
They thus expose new storage interfaces (a key-value store, a page server, a file service), as anticipated 
by Do et al.~\cite{do_programmable_2019} for on-path architectures, rather than the standardized computational storage interfaces. 
Second, they run into the same constraint as \delilah: memory on the device is the scarce resource. 
DDS keeps updates on the host for that reason, and Scalio has to keep DPU memory and SSD state consistent in software, 
since they are not kept coherent by hardware. 

Giouroukis et al.~\cite{Giouroukis2025-od} study how DPU accelerators perform on query-related tasks 
(compression, decompression and regular expression filtering) applied to data arriving from the network. 
In their setting, the host decides which tasks to offload and submits them one by one. 
They find that DMA-intensive offloads compete with the NIC for the DPU's memory interconnect, 
and that the best placement between host and DPU varies with the operator, the data and the query. 
Above all, they find that this fine-grained, host-driven offload pays off only when the accelerator configuration stays fixed: 
reconfiguring the device between tasks costs more than the processing itself, 
so that host CPUs outperform the DPU on small or variable-sized inputs. 
Our evaluation shows that per-invocation overhead also limits host-driven offload on computational storage: 
when the host invokes a function for each chunk of stored data, the cost of each invocation and of staging the chunk 
in device memory accounts for a large share of query execution. 
Rather than refining host-side offloading decisions, we argue that the iteration over data must move onto the device.

\subsection{Code Representation}
\label{sec:related:code}

Both standards distinguish between functions pre-installed on the device and functions downloaded by the host at run-time. 
In SNIA terms, functions execute in an engine environment (CSEE) such as an operating system, a container, an FPGA bitstream 
or an eBPF environment. NVMe distinguishes device-defined programs from downloadable programs, 
and proposes eBPF for the latter, on the grounds that it is a vendor-neutral instruction set architecture~\cite{bates_nvme_2022}. 
In the introduction, we argued that eBPF is best understood as a relocatable binary format, 
used to call device-defined functions rather than to express data processing. 
In SNIA terms, an eBPF program that calls registered functions is an aggregator CSF, running in an eBPF CSEE and invoking 
CSFs that may run in other CSEEs, e.g., in hardware. 
In this section, we review the characteristics of eBPF that support this view and that matter for computational storage.

\paragraph{eBPF.}
\label{sec:ebpf}
BPF stands for BSD (or Berkeley) Packet Filter. It was originally proposed in 1993 to 
perform user-defined packet filtering, e.g., in the context  of a firewall within the Unix kernel, without copying packets~\cite{mccanne_bsd_1993}. 
BPF defined a bytecode structure together with a virtual machine. 
In 2014, Alexei Starovoitov introduced extended BPF (eBPF), an adaptation of BPF for Linux and 
modern processors~\cite{starovoitov_lkml_nodate}\footnote{\url{https://lwn.net/Articles/598545/}}. 
eBPF is now used to execute user-defined functions inside the Linux kernel for performance monitoring~\cite{gregg_bpf_2019}, 
security monitoring~\cite{fournier_runtime_2021} and storage management~\cite{zhong_xrp_2022}, 
and as an intermediate representation for hardware-accelerated packet processing~\cite{brunella_hxdp_2022}.
The eBPF foundation was established in August 2021\footnote{The eBPF foundation (\url{http://ebpf.foundation}) is a Linux Foundation Project.} 
to organize the governance of this ecosystem, and the instruction set has since been 
standardized\footnote{\url{https://github.com/ebpffoundation/ebpf-docs}. 
}.

\paragraph{eBPF programs.}
\label{sec:programs}
An eBPF program is composed of a single main routine, located in memory in a {\em program slot} 
that can be read from the eBPF run-time environment.
It is a sequence of 64-bit encoded instructions, each composed of an 8-bit opcode, 
4-bit destination and source registers, a 16-bit offset and a 32-bit immediate value. 
Opcodes are grouped into classes (load, store, ALU, byteswap and branch). 
ALU, byteswap and load/store operations manipulate data at 64/32/16/8 bits granularity, 
and a wide 128-bit encoding appends a 64-bit immediate value to a basic instruction.
There are eleven registers: r0 holds the return value of a program, r1--r5 hold the arguments 
of called functions, r6--r9 are callee-saved, and r10 is a read-only frame pointer.
The call instruction uses its immediate value as an index in a function pointer table 
maintained by the eBPF run-time environment. 
The program counter is implicit: it cannot be manipulated by eBPF instructions and is managed by 
the run-time environment. As a consequence, eBPF does not support indirect jumps, indirect function calls
or jumps across eBPF programs\footnote{Within the Linux kernel, eBPF supports an additional operation: tail call.
It relies on an additional function pointer map to execute jumps to other eBPF programs 
(\url{https://blog.cloudflare.com/assembly-within-bpf-tail-calls-on-x86-and-arm}).}.
eBPF instructions do not support floating point operations.

\paragraph{Compilation and linking.}
Both gcc and clang/LLVM have an eBPF backend that generates eBPF programs from C programs\footnote{There is 
no support in eBPF for variadic functions or polymorphic types. As a result, C++ programs cannot be compiled into eBPF.}.  
Figure~\ref{lst:C} shows a simple C function that performs data massaging around a call to an external function. 
It takes a pointer as an argument and casts it as a struct pointer (line 6, the struct is defined in lines 1--4). 
It calls an external function (\texttt{regfunc}) on the string component of the input struct (line 7), 
and it writes, just beyond the input struct in memory, the product of the result of the external function 
with the integer element of the input struct (lines 8 and 9). 

\begin{figure}[htbp]
\centering
\begin{lstlisting}[style=ebpfC]
struct param {
    int n;
    char s[100];
};
int simple(void* ctx) {
    struct param *p = (struct param *) ctx;
    int v = regfunc(p->s);
    ctx += sizeof(struct param);
    * ((int*)ctx) = v*(p->n);
    return 0;
}
\end{lstlisting}
\caption{A C program calling an external function (simple.c).}
\label{lst:C}
\end{figure}

Figure~\ref{lst:eBPF} is the corresponding eBPF program in LLVM 
assembly\footnote{\url{https://releases.llvm.org/15.0.0/docs/CodeGenerator.html}}.
Register r1 contains the input pointer, which is first saved to the callee-saved register r6 (line 2),
then incremented by the size of the integer \texttt{n} (4 bytes) so that it points to the \texttt{s} component 
of the input struct (line 3). r1 is the argument of the external function call (line 4).
The integer value to which r6 points is then loaded into r1 (line 5), and multiplied by the return value 
of the external function, stored in r0 (line 6).
r1 is stored at the first memory address beyond the input struct, i.e., at r6 + 104 (line 7). 
Finally, the return value 0 is assigned to r0 (line 8) and the program exits (line 9).

\begin{figure}[htbp]
\centering
\begin{lstlisting}[style=ebpfasm] 
simple:
    r6 = r1
    r1 += 4
    call regfunc
    r1 = *(u32 *)(r6 + 0)
    r1 *= r0
    *(u32 *)(r6 + 104) = r1
    r0 = 0
    exit
\end{lstlisting}
\caption{The eBPF program generated from simple.c in LLVM assembly (clang -O1 -S -target bpf simple.c).}
\label{lst:eBPF}
\end{figure}

Note that \texttt{regfunc} is not resolved at compile time: the bytecode refers to it by name.
Note also that the compiler does not take into account the characteristics of the eBPF run-time environment, 
which might only allow programs of a given size or with predefined properties; 
for instance, the compiler can generate eBPF programs that contain loops.
We can distinguish the following phases in the life cycle of an eBPF program:
\begin{enumerate}
  \item {\em Produce bytecode} by compiling a C program;
  \item {\em Link bytecode} by replacing references to external functions with their
    index in the function pointer map of the run-time environment, or offsets into structs with their 
    actual offset in memory;
  \item {\em Load bytecode} into the run-time environment; 
  \item {\em Verify bytecode}, possibly rewriting it to be safe;
  \item {\em Execute bytecode} with a virtual machine, or as native code if it has been JIT-compiled in phase 3 or 4. 
\end{enumerate}
The linking phase is what makes eBPF a relocatable binary format. 
The code of external functions is linked with the process that runs the run-time environment. 
When an external function is registered, its name and function pointer are added to the external function pointer map, 
and a lookup by name returns its index in this map. 
The same bytecode can thus call a function implemented in software on one device and hardware-accelerated on another, 
as long as both devices register it under the same name.

\paragraph{Run-time environments.}
\label{sec:runtime}
There exist several software implementations of the eBPF ISA. 
The most prominent are the eBPF subsystem within the Linux kernel, which consists of an interpreter and 
several JIT compilers along with a run-time environment, and uBPF, a library that implements a virtual machine 
with a JIT compiler, designed to run in user-space and to be embedded in kernel-space\footnote{uBPF 
is the core component of the eBPF for Windows project (\url{https://github.com/microsoft/ebpf-for-windows}).
rBPF is a similar implementation in Rust (\url{https://github.com/qmonnet/rbpf}).}.
Hardware implementations of the eBPF ISA have also been proposed, most prominently 
hBPF\footnote{\url{https://github.com/rprinz08/hBPF}} and 
Sephirot, the eBPF processor of hXDP~\cite{brunella_hxdp_2022}, a design specialized for packet processing on FPGA NICs. 
A tool is available to evaluate the conformance of an eBPF run-time environment to the 
instruction set\footnote{\url{https://github.com/Alan-Jowett/bpf_conformance/}}. 

\paragraph{Memory.}
\label{sec:memory}
The stack and the program slot are regions of memory associated with the eBPF run-time environment. 
The size of the stack and the maximum number of instructions in the program slot are parameters of the environment. 
In uBPF, they are set by default to 512~B and 64K instructions respectively.
In the Linux kernel, eBPF programs can exchange data with user-space, or maintain state across executions, 
through key-value data structures denoted maps. 
More importantly for us, the caller of an eBPF program usually passes a memory buffer as a {\em context}. 
By convention, the first argument of an eBPF program is a pointer to this context (as in Figure~\ref{lst:C}). 
Most eBPF programs in the Linux kernel (e.g., socket filters, tc filters, XDP programs) and all uBPF programs 
respect this convention. A socket filter, for instance, takes a pointer to an \texttt{\_\_sk\_buff}, 
whose \texttt{data\_end} field points to the end of the context so that its length can be computed. 
The context is the natural vehicle for arguments and results when eBPF programs are offloaded to computational storage: 
it maps onto function data memory in SNIA and subsystem local memory in NVMe.

\paragraph{Verification.}
The eBPF verifier in Linux enforces that eBPF programs terminate without using too many resources 
and without leaking the contents of kernel memory. It is now over 15K lines of 
code\footnote{\url{https://github.com/torvalds/linux/blob/master/kernel/bpf/verifier.c}}. 
In contrast, the uBPF loader merely checks that an eBPF program fits inside the allocated program slot.
The eBPF for Windows project uses the PREVAIL verifier~\cite{gershuni_simple_2019} together with uBPF.
Verification constraints designed for the kernel, such as bounded loops, 
are not necessarily appropriate for computational storage.

\paragraph{eBPF as a data processing language.}
Huang et al.~\cite{huang_evaluation_nodate} compare eBPF and WebAssembly as code representations for computational storage. 
They argue that the negative aspects of eBPF (no support for floating point operations, no support for 64-bit division) 
outweigh its positive aspects (low overhead, both in terms of bytecode footprint and execution time). 
Their comparison rests on the hypothesis that the offloaded bytecode performs the data processing itself, 
and it does not distinguish between device-defined functions and downloadable programs. 
We take a different position: eBPF programs are glue that massages 
the inputs and outputs of registered functions and orchestrates their execution, 
while the bulk of the processing is performed by native or hardware-accelerated code. 
Whatever its representation, however, a downloaded program remains a host-invoked function operating on data 
moved by the host. The code representation does not change the execution model.

\section{\delilah}
\label{sec:delilah}

\delilah is a computational storage system that implements the execution model of the
standard computational storage interfaces on an actual open-hardware device.
It uses eBPF to represent the functions offloaded by the host.
A first version of \delilah was presented in~\cite{hedam_delilah_2023}. Here, we detail its design
and implementation; Appendix~\ref{app:overview} details its software and hardware components and its memory layout.

\subsection{Design}
\label{sec:delilah:design}
\label{sec:delilah:decisions}
\label{sec:delilah:requirements}
\label{sec:delilah:arch}
\label{sec:designspace}

\begin{figure}[tbp]
  \centering
  \definecolor{dsUser}{HTML}{BBBBFF}
\definecolor{dsDelilah}{HTML}{00EE00}
\definecolor{dsSys}{HTML}{999999}
\definecolor{dsRam}{HTML}{FF8080}
\begin{tikzpicture}[
  x=0.0061cm, y=-0.0061cm,
  font=\footnotesize,
  every node/.style={align=center, inner sep=1pt},
  usr/.style={draw=black, line width=0.6pt, fill=dsUser},
  dlh/.style={draw=black, line width=0.6pt, fill=dsDelilah},
  sys/.style={draw=black, line width=0.6pt, fill=dsSys},
  ram/.style={draw=black, line width=0.6pt, fill=dsRam},
  ssd/.style={draw=black, line width=0.6pt, fill=black},
  wire/.style={line width=0.6pt},
]
\newcommand{\bx}[5]{%
  \draw[#3] (#1) rectangle (#2);
  \node[text=#4] at ($(#1)!0.5!(#2)$) {#5};}

\draw[line width=0.6pt] (212,4) rectangle (1225,472);
\draw[line width=0.6pt] (212,175) -- (1225,175);
\node[anchor=south east] at (1220,172) {user space};
\node[anchor=north east] at (1222,180) {kernel space};

\draw[wire] (674,137) -- (674,205);
\draw[wire] (588,301) -- (537,331);
\draw[wire] (770,301) -- (825,331);

\bx{435,40}{912,137}{usr}{black}{User Program}
\bx{428,205}{920,301}{dlh}{black}{Delilah Driver}
\bx{249,331}{652,428}{sys}{white}{XDMA Driver}
\bx{740,331}{1102,428}{sys}{white}{PCIe Driver}

\draw[{Latex[length=4pt]}-{Latex[length=4pt]}, line width=0.8pt] (718,478) -- (718,566);

\draw[line width=0.6pt] (4,566) rectangle (1262,1593);
\node[anchor=north west] at (14,580) {PL};

\shade[top color=white, bottom color=black!12] (4,797) rectangle (558,1342);
\shade[top color=white, bottom color=black!12] (558,797) rectangle (785,1342);
\draw[line width=0.6pt] (4,797) rectangle (785,1342);
\draw[line width=0.6pt] (558,797) -- (558,1342);
\node[anchor=north west] at (14,805) {PS};
\node[anchor=north west, align=left] at (566,800) {kernel\\space};
\node[anchor=south east] at (520,1335) {user space};

\draw[wire] (633,730) -- (588,797);
\draw[wire] (745,730) -- (1037,1088);
\draw[wire] (897,718) -- (942,728);
\draw[wire] (963,834) -- (785,922);
\draw[wire] (1055,834) -- (738,1430);
\draw[wire] (785,1088) -- (942,1118);
\draw[wire] (1090,1214) -- (1090,1430);
\draw[wire] (785,1308) -- (987,1430);
\draw[wire] (595,1342) -- (655,1430);
\draw[wire] (290,953) -- (275,998);
\draw[wire] (395,953) -- (571,978);
\draw[wire] (458,1178) -- (571,1210);
\draw[wire] (292,1233) -- (236,1363);
\draw[wire] (1240,768) -- (1264,768);
\draw[wire] (1240,1154) -- (1264,1154);
\draw[wire] (703,1557) -- (701,1698);
\draw[wire] (1090,1557) -- (1093,1698);

\bx{473,588}{897,730}{sys}{white}{DMA Engine\\over PCIe}
\bx{942,708}{1240,834}{sys}{white}{Memory\\Interface}
\bx{942,1088}{1240,1214}{sys}{white}{Memory\\Interface}
\bx{1264,611}{1352,923}{ram}{black}{\rotatebox{90}{RAM}}
\bx{1264,998}{1352,1310}{ram}{black}{\rotatebox{90}{RAM}}

\bx{63,857}{555,953}{dlh}{black}{Delilah Controller}
\bx{56,998}{458,1095}{sys}{white}{UBPF VM}
\bx{190,1095}{458,1233}{usr}{black}{Registered\\Functions}
\bx{571,924}{763,1061}{sys}{white}{udma\\buf}
\bx{571,1173}{763,1311}{sys}{white}{NVMe\\Driver}
\bx{56,1363}{324,1564}{usr}{black}{Hardware\\Accelerated\\Functions}

\bx{555,1430}{853,1557}{sys}{white}{PCIe Bridge}
\bx{942,1430}{1240,1557}{sys}{white}{PCIe Bridge}

\bx{592,1698}{808,1998}{ssd}{white}{SSD}
\bx{987,1698}{1203,1998}{ssd}{white}{SSD}
\end{tikzpicture}
  \caption{\delilah overview. \label{fig:archi}}
\end{figure}

\paragraph{Platform.}
We use OpenSSD Daisy~\cite{daisy-sk}
as hardware platform for our computational storage device. Daisy is the latest generation of OpenSSD
prototypes, designed by Prof. Song and his team at Hanyang University,
and commercially available through CRZ Technology.
It is an accelerator card equipped with 2x100GE and PCIe Gen3x16 connectors,
a Zynq Ultrascale+ Multiprocessor System on a Chip (MPSoC)
and a backplane interface for connecting two M.2 SSDs.
The Zynq Ultrascale+ MPSoC combines an FPGA, denoted PL (programmable logic), with four ARM Cortex-A53 cores
running embedded Linux, denoted PS (processing system).
The PS only has access to peripherals (PCIe, RAM DIMMs, M.2 SSDs) through the PL.
The boot image of Daisy therefore contains a PL image, i.e., an FPGA bitstream synthesized with the Vivado toolchain
from a block design that combines generic IP (e.g., a DMA engine) with accelerators defined in a hardware description
language, and a PS image, i.e., embedded Linux with programs cross-compiled for the ARM cores.

We use Daisy in an on-path PCIe architecture:
Daisy is a PCIe device for the host and a PCIe root for the SSDs.
A computational storage device accesses SSDs through their external interface.
In an on-path architecture, the host thus reaches stored data through at least two DMA copies, one between host and device
and one between device and SSD; in an off-path architecture, processing on the device requires the same number of
copies, performed with peer-to-peer DMA. In both cases, all copies go through device memory.

\paragraph{Components.}
The \delilah software consists of two components: the \delilah driver, a kernel module on the host, and the
\delilah controller, a user-space program on the device CPU (PS).
The driver exposes a character device, through which applications submit commands with io\_uring; it transfers data by
DMA between host memory and device memory, and passes commands to the controller through a memory region
mapped in the host address space (BAR0).
The controller runs a number of eBPF virtual machines, each executing an eBPF program on behalf of the host.

eBPF is the means by which the host calls device functions.
A computational storage device comes with device-specific functions, implemented in software or in hardware.
One option is to call each of them with a dedicated NVMe command (e.g.,~\cite{do_better_2021}); NVMe is extensible,
but every new function then requires extending the NVMe controller.
\delilah instead uses eBPF as a relocatable binary format.
Device-specific functions are linked and registered with the eBPF run-time environment on the device;
we denote them {\em registered functions}. To call them, the host generates an eBPF program that references them by
name, and these references are resolved when the program is linked on the device.
The host can thus call any registered function without extending the host-device interface.

Several eBPF run-time environments could host these programs: the Linux kernel run-time environment, uBPF in user-space,
and hardware eBPF processors. The Linux kernel run-time environment is the most robust and mature,
and might suit a smart NIC where data flows through the network and storage stacks inside the kernel; hardware eBPF
processing might suit FPGA accelerator cards, especially in off-path architectures.
The \delilah controller runs uBPF virtual machines, which give us the most flexibility to explore the design space:
registered functions are ordinary user-space functions on the PS, which access the SSDs through the NVMe driver of
the embedded Linux, and can in turn call hardware accelerators on the PL.
Put differently, eBPF programs massage the inputs and outputs of registered functions and orchestrate their execution, as in
Figure~\ref{lst:C}, while the bulk of the processing is performed by registered functions, possibly hardware
accelerated, e.g., to convert Apache Parquet to Arrow in FPGA~\cite{peltenburg_battling_2020}.

Arguments and results are exchanged through the abstraction that the eBPF run-time environment already provides:
the memory buffer passed to a program as its context (Section~\ref{sec:memory}), which we denote the {\em data slot}.
\delilah exposes data slots, as well as the {\em program slots} that hold eBPF programs, as memory shared between
host and device. The host writes a program into a program slot and its arguments into a data slot,
requests the execution of the program on the data slot, and reads the results back from the data slot.
All program and data slots have the same size, determined on the device and reported to the host through two 4-byte fields.
In theory, these slots have a maximum size of 4GB each, but in practice, their size is determined by the amount
of physical memory on the device that is dedicated to data transfer.

Code offloaded from the host must not be able to exploit the eBPF run-time environment to corrupt or deny access to
stored data. This requires verifying the eBPF virtual machine or JIT compiler~\cite{yuan_end--end_2022,nelson_specification_2020},
as well as eBPF programs and the registered functions
they call. The Linux eBPF verifier does not suit computational storage: backward jumps are essential for data massaging,
which makes termination analysis much harder, while termination remains necessary so that the device is not locked up
by an infinite loop. Verifying offloaded code for computational storage is thus an open problem, which starts with
specifying appropriate safety and security policies.
We leave it out of scope: like Eid-Hermes~\cite{oliveira_eid-hermes_2022}, \delilah relies on the uBPF loader,
which only checks that a program fits in its program slot.

\paragraph{Block design and device memory.}
The {\em block design} on the PL connects the PS to the outside world: a DMA engine over PCIe for
communication with the host, memory interfaces for accessing RAM, and PCIe bridges for communicating with the M.2 SSDs.
The block design used in our experiments, built from Xilinx IP modules connected through AXI interconnects
(Appendix~\ref{app:hardware}), connects the device to the host over PCIe Gen3 x8 (8~GB/s), and an M.2 SSD over
PCIe Gen3 x4 (4~GB/s); a DDR4 DIMM attached to the PL holds the program and data slots.

Each DDR4 memory channel provides a peak bandwidth of 12.8 GB/s. Consider a function executed on
data read from an SSD. The data is first written to device memory, then read by the PS
via load instructions; the PS writes its result back to device memory via store instructions, and the
host subsequently reads the result from device memory. Can the device process SSD data at 4 GB/s?
The answer is no: each byte of input generates four memory accesses (two writes and two reads),
requiring 16 GB/s of memory bandwidth, which exceeds the channel’s 12.8 GB/s peak, and practical
sustained bandwidth is lower still.
The memory access pattern is thus a bottleneck.

This pattern limits end-to-end throughput to 12.8 / 4 = 3.2~GB/s per DIMM;
with the single DIMM used in our experiments, this is the bound for the whole device, well below the 8~GB/s of the host link.
Again, it is not memory bandwidth, but the way compute is attached to memory and the way data is staged in
memory that constitute the bottleneck.

\delilah stages data retrieved from storage in a {\em shared buffer}, a region of device memory, distinct from
the program and data slots, that is accessible to all virtual machines.

We detail how to process queries with such a computational storage platform in the next Section.
Before we get there, let us focus on the \delilah{} implementation.

\subsection{Implementation}
\label{sec:delilah:impl}

We implemented the block design with Vivado 2019.1.
The \delilah controller and the registered functions are programmed in C and cross-compiled with gcc 12.1
for PetaLinux 2019.1, the embedded operating system running on the PS.
The \delilah driver is also implemented in C and compiled with gcc 12.1.
Appendix~\ref{app:overview} details the software components (Appendix~\ref{app:software}), the hardware
components (Appendix~\ref{app:hardware}) and the memory layout (Appendix~\ref{app:memory}), with pointers to the code.

\subsubsection{Transport Protocol}
\label{sec:delilah:protocol}
\label{sec:delilah:driver}
Driver and controller implement a transport protocol between host and device,
which is a simplified form of the memory, copy and compute requests of the SNIA and NVMe APIs.
Unlike the reference flow of the API, however, data is loaded from the SSDs into device memory by registered functions
rather than by host-issued storage requests.

The protocol defines four commands for eBPF programs:
(i) transfer an eBPF program from the host to a program slot,
(ii) write program arguments to a data slot, (iii) execute an eBPF program using a program and a data slot, and (iv) read results from a data slot.
The current version of \delilah supports a single command
for executing an eBPF program. The rationale for defining a single command, instead of separate commands for
loading a program and executing it, is to avoid a roundtrip when executing a program once.
Defining separate commands for JIT compilation followed by multiple program executions with
different arguments is future work.
In addition to these four commands, the protocol defines three commands for
clearing the device caches, clearing the controller state and
returning device information.

Programs, arguments and results are transferred by DMA between host memory and the slots.
The actual address of each slot is computed by the host from an offset contained in BAR0 (see Figure~\ref{fig:memlayout},
this offset is configured by the Delilah controller when it starts).
To execute a program, the driver writes the command to BAR0 and sets a control bit.
The controller spawns one thread per uBPF virtual machine, pinned to its own PS core; each thread polls the control
register of its virtual machine, loads the program from the program slot named in the command, and executes it on the
data slot named in the command. Any virtual machine can thus execute any program slot on any data slot.
On completion, the controller writes the response to BAR0 and signals the host with an MSI-X interrupt.

\subsubsection{DMA Buffers and Cache Coherence}
\label{sec:delilah:coherence}
Program and data slots are DMA buffers that the controller maps into user space. They reside in the PL DDR4,
which the host and the SSD access by DMA without snooping the caches of the device processors.
Coherence must thus be maintained in software, by invalidating the slots before a program executes and
flushing the data slot afterwards.

This leads to a trade-off.
With {\em whole-slot maintenance}, the protocol is simple, but the controller, which does not know what the host
exchanges, maintains entire slots, at a cost proportional to their size that dominates the latency of small
programs (Section~\ref{sec:eval:coherence}).
With {\em range-based maintenance}, the execution command carries the ranges of the data slot that the host has
written and will read back: the cost becomes proportional to the data exchanged, but coherence becomes part of the
host-device protocol, and the host must declare its ranges correctly, at cache-line granularity.
With {\em hardware coherence}, the slots would reside in the PS DDR and all DMA transfers would go through ports
that the CCI-400 interconnect keeps coherent with the processor caches: neither maintenance nor declarations would
be needed, but the slots would be confined to the 2~GB of PS DDR shared with Linux, and transfers would compete for
the coherent ports.
Hardware coherence thus trades protocol complexity for device memory capacity and bandwidth, the scarce resources
on the device (Section~\ref{sec:delilah:design}); we have not implemented it.

\subsubsection{Hardware-Accelerated Filtering}
\label{sec:delilah:hwfilter}

Registered functions can also call accelerators in the PL. As a first case, we integrated a filter synthesized
with Vivado HLS~\cite{hedam_delilah_thesis_2024}, which supports five predicates (equal, not equal, less or equal,
greater or equal, and between) and writes a position list; a later revision of the block design adds four instances
of this filter (Appendix~\ref{app:hardware}).
In the controller, five registered functions, one per predicate, are a facade for a single function that drives the
accelerator. Because the accelerator accesses memory directly, bypassing the caches
of the device processors, this function has to do what the uBPF virtual machine does not: it translates the pointers
of the eBPF program, which are relative to the data slot or the shared buffer, into physical addresses, and it
flushes the input range before starting the accelerator and invalidates the output range after completion.
Integrating an accelerator thus turns the coherence problem
between host and device (Section~\ref{sec:delilah:coherence}) into a coherence problem between the device processors
and the PL. We evaluate the accelerator in Section~\ref{sec:eval:qp}.

\subsubsection{Block Design Synthesis}
\label{sec:delilah:synthesis}

The original version of our block design, with a 16-lane PCIe link to the host,
cannot be synthesized for the Daisy FPGA because of routing congestion;
reducing the number of PCIe lanes from 16 to 8 results in successful and much faster synthesis
(Appendix~\ref{app:hardware}).
This points to a trade-off: any specialized block that reduces bandwidth facilitates routing without sacrificing
end-to-end performance, as long as the reduced bandwidth remains above the memory bottleneck, and a block design can
be optimized for throughput (with PL DIMMs) or latency (with the PS memory only).
The block design should thus be considered use case specific.

\section{Offloaded Query Processing}
\label{sec:qp}

We hardcode a query and focus on the eBPF programs that are downloaded to the device: this is a sanity check 
before we invest in integrating \delilah with the storage engine of a database system.
To offload query processing, we implement relational operators 
(e.g., filters, conversions and aggregations) as eBPF programs, 
which may delegate their inner loop to registered functions on the device. 
Filters, for instance, invoke a registered function vectorized with ARM Neon SIMD instructions, 
passing it the parameters of the operator (e.g., the predicate constants) read from the data slot.

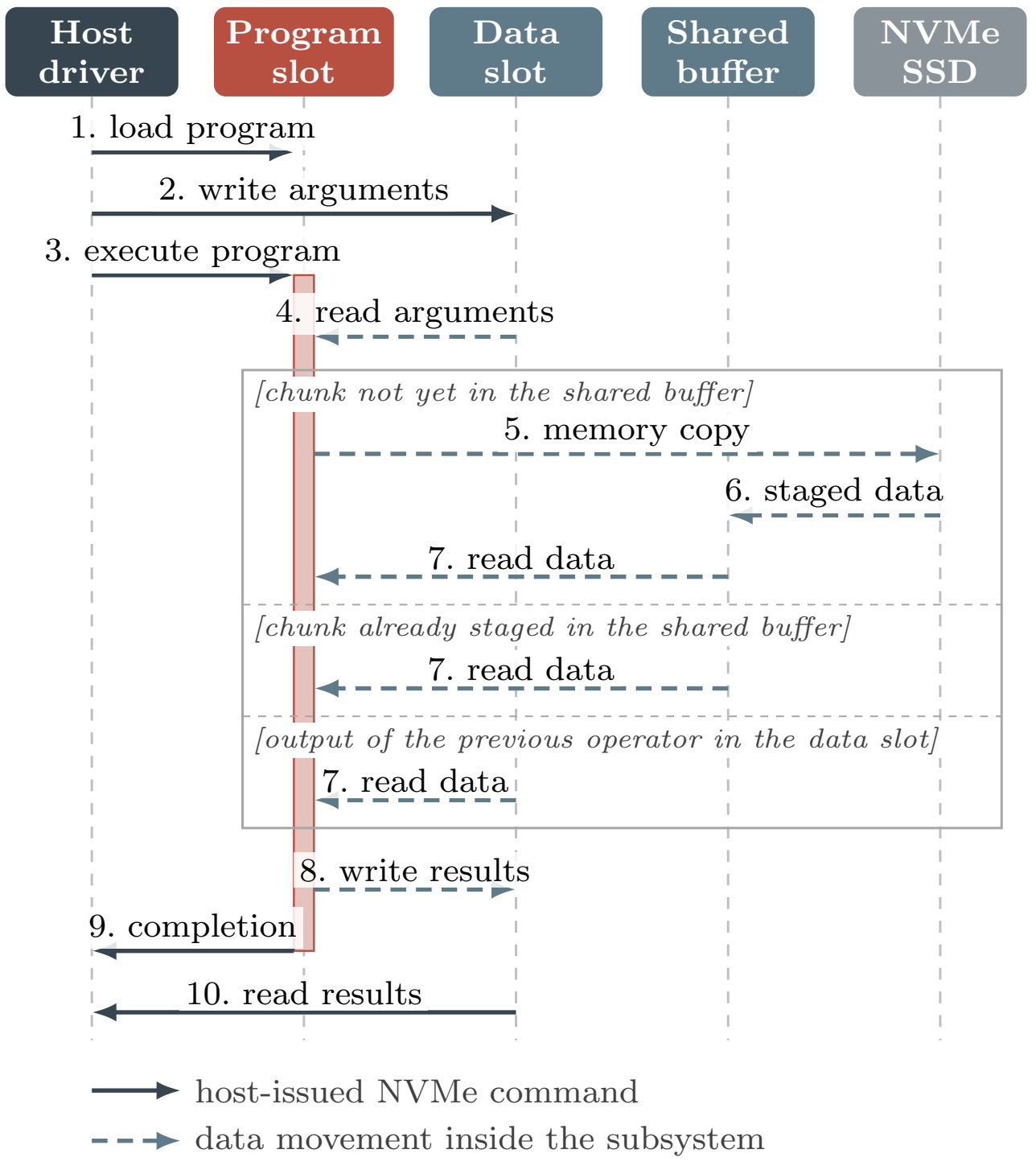
\begin{figure}[htbp]
  \definecolor{seqHost}{HTML}{36454F}
\definecolor{seqProg}{HTML}{B85042}
\definecolor{seqSlot}{HTML}{5F7A88}
\definecolor{seqSSD}{HTML}{8A9399}
\resizebox{\linewidth}{!}{%
\begin{tikzpicture}[
  font=\footnotesize,
  head/.style={rounded corners=3pt, minimum width=1.55cm, minimum height=0.8cm, align=center,
               text=white, font=\footnotesize\bfseries},
  life/.style={draw=gray!50, dashed, line width=0.6pt},
  cmd/.style={draw=seqHost, line width=1.1pt, -{Latex[length=2.2mm]}},
  mov/.style={draw=seqSlot, line width=1.1pt, dash pattern=on 4pt off 2.5pt, -{Latex[length=2.2mm]}},
  lbl/.style={above, inner sep=1.5pt, fill=white, fill opacity=0.85, text opacity=1},
  guard/.style={anchor=north west, font=\scriptsize\itshape, text=black!75, inner sep=1pt, fill=white},
  sep/.style={draw=gray!70, dash pattern=on 2pt off 2pt},
]
\def\xH{0} \def\xP{1.9} \def\xD{3.8} \def\xS{5.7} \def\xN{7.6}
\def\xL{1.35} \def\xR{8.15}  %
\def\yEnd{-8.85}

\node[head, fill=seqHost] at (\xH,0) {Host\\driver};
\node[head, fill=seqProg] at (\xP,0) {Program\\slot};
\node[head, fill=seqSlot] at (\xD,0) {Data\\slot};
\node[head, fill=seqSlot] at (\xS,0) {Shared\\buffer};
\node[head, fill=seqSSD]  at (\xN,0) {NVMe\\SSD};
\foreach \x in {\xH,\xP,\xD,\xS,\xN} \draw[life] (\x,-0.4) -- (\x,\yEnd);

\fill[seqProg!35] (\xP-0.09,-2.0) rectangle (\xP+0.09,-8.05);
\draw[seqProg, line width=0.5pt] (\xP-0.09,-2.0) rectangle (\xP+0.09,-8.05);

\draw[cmd] (\xH,-0.9)  -- node[lbl] {1. load program}    (\xP-0.09,-0.9);
\draw[cmd] (\xH,-1.45) -- node[lbl] {2. write arguments} (\xD,-1.45);
\draw[cmd] (\xH,-2.0)  -- node[lbl] {3. execute program} (\xP-0.09,-2.0);
\draw[mov] (\xD,-2.55) -- node[lbl] {4. read arguments}  (\xP+0.09,-2.55);

\draw[gray!70, line width=0.6pt] (\xL,-2.85) rectangle (\xR,-6.95);
\node[guard] at (\xL+0.05,-2.9) {[chunk not yet in the shared buffer]};
\draw[mov] (\xP+0.09,-3.6) -- node[lbl] {5. memory copy} (\xN,-3.6);
\draw[mov] (\xN,-4.15)     -- node[lbl] {6. staged data} (\xS,-4.15);
\draw[mov] (\xS,-4.7)      -- node[lbl] {7. read data}   (\xP+0.09,-4.7);

\draw[sep] (\xL,-4.95) -- (\xR,-4.95);
\node[guard] at (\xL+0.05,-5.0) {[chunk already staged in the shared buffer]};
\draw[mov] (\xS,-5.7) -- node[lbl] {7. read data} (\xP+0.09,-5.7);

\draw[sep] (\xL,-5.95) -- (\xR,-5.95);
\node[guard] at (\xL+0.05,-6.0) {[output of the previous operator in the data slot]};
\draw[mov] (\xD,-6.7) -- node[lbl] {7. read data} (\xP+0.09,-6.7);

\draw[mov] (\xP+0.09,-7.5)  -- node[lbl] {8. write results} (\xD,-7.5);
\draw[cmd] (\xP-0.09,-8.05) -- node[lbl] {9. completion}    (\xH,-8.05);
\draw[cmd] (\xD,-8.6)       -- node[lbl] {10. read results} (\xH,-8.6);

\draw[cmd] (0,-9.3)  -- (0.8,-9.3)  node[right, text=black!75] {host-issued NVMe command};
\draw[mov] (0,-9.75) -- (0.8,-9.75) node[right, text=black!75] {data movement inside the subsystem};
\end{tikzpicture}%
}
  \caption{Chunked execution of an offloaded operator over a column $X$ split into chunks $C_0 \ldots C_n$: 
  interactions for one chunk. Steps 5--7 depend on where the chunk is (the three alternatives are shown in the frame). 
  Solid arrows are host-issued commands; dashed arrows are data movements inside the storage subsystem.}
  \label{fig:chunked}
\end{figure}

The columns of the base tables are stored on the SSD, in binary format. Because device memory is the scarce 
resource (Section~\ref{sec:delilah:design}), a column cannot in general be loaded into device memory at once. 
We therefore split each column into chunks that fit into a {\em shared buffer}, 
a region of device memory, distinct from the program and data slots, into which data is staged from the SSD.
Figure~\ref{fig:chunked} shows the sequence of interactions for processing one chunk:
\begin{enumerate}
    \item the host driver loads the eBPF program into a program slot;
    \item the host writes the program arguments (e.g., the column and the chunk to process, 
    as well as the state carried over from previous chunks) to the data slot;
    \item the host issues the execute command;
    \item the eBPF program reads its arguments from the data slot;
    \item if the chunk is not yet in the shared buffer, a registered function reads it from the NVMe SSD 
    into the shared buffer, from which the operators of the current pass read it without accessing the SSD again; 
    since a column is in general much larger than the shared buffer, it is staged chunk by chunk once per pass; 
    the step is skipped when the chunk is already in the shared buffer, 
    or when the operator consumes the output that the previous operator left in the data slot;
    \item the chunk is staged in the shared buffer;
    \item the operator reads its input: the chunk staged in the shared buffer, 
    or the output of the previous operator in the data slot;
    \item the operator writes its results to the data slot;
    \item the device signals completion to the host;
    \item the host reads the results from the data slot.
\end{enumerate}
Steps 1--3, 9 and 10 are host-issued commands, while steps 4--8 correspond to data movement inside the 
storage subsystem.
To overlap processing on the device with transfers to the host, two chunks are kept in flight: 
while the device processes chunk $C_{i+1}$, the host fetches the results of chunk $C_i$ 
(we refer to this pattern as zig-zag).

Listings~\ref{lst:filter}--\ref{lst:convert} show sanitized versions of two of the operators used for 
SSB Q3.3 (Section~\ref{sec:eval:qp}). 
Both follow the steps above: the eBPF program reads its arguments from the data slot (step~4), 
reads its input, either the chunk staged in the shared buffer or the output of the previous operator 
in the data slot (step~7), and writes its results to the data slot, after its arguments (step~8). 
The two operators differ in where their loop runs. 
The filter (Figure~\ref{lst:filter}) is a thin eBPF program that delegates the loop to a registered function, 
compiled natively and vectorized with Neon (Figure~\ref{lst:neon}). 
Each iteration of the inner loop compares four 32-bit values with the predicate constants and sets one bit per value; 
the loop has a fixed trip count of 32, so that a batch of 128 values yields four 32-bit bitmask words, 
and the native compiler can unroll it completely. 
The conversion (Figure~\ref{lst:convert}), in contrast, runs its loop in eBPF. 
The eBPF program is compiled ahead of time by clang and translated instruction by instruction by the uBPF JIT, 
which does not optimize the code, so the loop must be tuned at the source level: 
the branch on the conversion type is hoisted out of the loop, and the conversion of the four digits 
of the year is unrolled by hand. 
Both operators access their input sequentially; we discuss the cost of memory accesses 
from the device processors in Section~\ref{sec:eval:overhead}.

\begin{figure}[htbp]
\begin{lstlisting}[style=ebpfC]
int filter(void *ctx, int size, void *buf,
           int buf_size) {
  filter_op *op = (filter_op *)ctx;
  uint32_t n = op->file.size / sizeof(uint32_t);
  /* results follow the arguments in ctx */
  uint32_t *out = ctx + op->padding;
  /* input: output of the previous operator,
     or chunk staged in the shared buffer */
  uint32_t *in = op->reuse_data
               ? out : buf + op->cached_offset;
  /* the loop runs in a registered function */
  if (op->comp_type == EQ)
    op->count = filter_eq_neon(out, in, n,
                               op->comp0);
  else /* BETWEEN */
    op->count = filter_bw_neon(out, in, n,
                    op->comp0, op->comp1);
  return 0;
}
\end{lstlisting}
\caption{eBPF program for the filter operator (sanitized).}
\label{lst:filter}
\end{figure}

\begin{figure}[htbp]
\begin{lstlisting}[style=nativeC]
/* registered function, compiled natively */
uint32_t filter_bw_neon(uint32_t *out,
    const uint32_t *in, uint32_t n,
    uint32_t lo, uint32_t hi) {
  uint32x4_t vlo = vdupq_n_u32(lo);
  uint32x4_t vhi = vdupq_n_u32(hi);
  uint32_t *start = out;
  const uint32_t *end = in + n - n % 128;
  while (in != end) {  /* 128 values per batch */
    uint32x4_t mask = vdupq_n_u32(0);
    uint32x4_t bit = vdupq_n_u32(1);
    /* fixed trip count (32): fully unrollable */
    for (int i = 0; i < 32; i++, in += 4) {
      uint32x4_t v = vld1q_u32(in);
      uint32x4_t m = vandq_u32(vcgeq_u32(v, vlo),
                               vcleq_u32(v, vhi));
      mask = vorrq_u32(mask, vandq_u32(bit, m));
      bit = vshlq_n_u32(bit, 1);
    }
    vst1q_u32(out, mask);  /* 4 x 32-bit masks */
    out += 4;
  }
  /* remaining n % 128 values: scalar loop
     with #pragma unroll(8), omitted */
  return (char *)out - (char *)start;
}
\end{lstlisting}
\caption{Registered function implementing the filter loop with Neon intrinsics (sanitized; the equality 
filter differs only in the comparison, lines~15--16).}
\label{lst:neon}
\end{figure}

\begin{figure}[htbp]
\begin{lstlisting}[style=ebpfC]
int convert_year(void *ctx, int size, void *buf,
                 int buf_size) {
  conversion_op *op = (conversion_op *)ctx;
  uint32_t *out = ctx + op->padding;
  /* input: chunk staged in the shared buffer */
  char *in = buf + op->cached_offset;
  char *end = in + op->file.size;
  /* the loop runs in eBPF: dates are 8 chars
     (YYYYMMDD); the branch on the conversion type
     is hoisted out of the loop (other types
     omitted); the digits are unrolled */
  for (; in != end; in += 8)
    *out++ = (in[0] - '0') * 1000
           + (in[1] - '0') * 100
           + (in[2] - '0') * 10
           + (in[3] - '0');
  op->count = out - (uint32_t *)
              (ctx + op->padding);
  return 0;
}
\end{lstlisting}
\caption{eBPF program for the conversion of dates to years (sanitized).}
\label{lst:convert}
\end{figure}

This execution model has a clear strength: each operator implements a loop over a single chunk, 
and this loop is tight on the device. However, it inherits the host-centric execution model of the 
standard computational storage interfaces (Section~\ref{sec:related:execution}), with three consequences:
\begin{itemize}
    \item The host manages the loop across chunks. Every chunk requires a round of host-device 
    commands, which introduces synchronization overhead.
    \item State that must persist across chunks (e.g., partial aggregates or the positions selected by a filter) 
    is carried as program arguments through the data slot, which may lead to pointer chasing on the device.
    \item A query is executed as a sequence of loops, one per operator, each invoked from the host. 
    Processing is thus column-at-a-time, in chunks, and intermediate results are materialized 
    between operators rather than pipelined on the device.
\end{itemize}

Could the device drive the loop?
One could try to avoid host-driven iteration by fusing the operators of a query into a single eBPF program
that loops over all chunks on the device. Within the standard execution model, this does not work.
On the input side, the program would have to read the next chunk from the SSD itself. The standard interfaces
leave this to the host: SNIA storage requests and NVMe copy commands are host-issued, so a device that follows
the standards strictly returns to the host for every chunk. Only the deviation of \delilah, registered functions that
read from the SSD, makes device-side iteration over the input possible.

On the output side, the results of a program reach the host only through its data slot: the shared buffer is not
accessible to the host, and the execution command completes only when the program terminates, so the device cannot
stream partial results to the host while it keeps running. When the output of a fused program exceeds its data slot,
the program must stop, the host must drain the slot, and the host must invoke the program again, with the state of
the computation carried through the data slot. Iteration thus returns to the host, driven by the size of the output
rather than by the size of the input. Larger data slots only push this limit back: they are bounded by device memory,
and with whole-slot maintenance the cost of coherence grows with their size (Section~\ref{sec:eval:coherence}).
The final result of SSB Q3.3 is small, since it is grouped on a few cities and years, but intermediate results that
the host needs, e.g., to execute joins, and the results of less selective queries, are not.

We evaluate the consequences of this execution model on the Star Schema Benchmark in Section~\ref{sec:eval:qp}.

\section{Evaluation}
\label{sec:evaluation}

We use \delilah to answer the questions raised by its design (Section~\ref{sec:delilah:design}). 
The first three concern the mechanisms on which the host-orchestrated execution model rests: 
(a)~what does it cost to invoke an eBPF program, and the registered functions it calls, on the device? 
(b)~what does it cost to keep the memory shared between host and device coherent? 
(c)~how efficiently can the host read and write device memory? 
The fourth question is the one this paper addresses: 
(d)~is computational storage based on host-invoked functions and host-managed data movement a good match
for query processing, and does a single hand-coded query suffice to tell?

\subsection{Experimental Framework}
\label{sec:eval:framework}

We focus on latency, i.e., the time it takes to transfer data to and from the device and to execute 
eBPF programs on it. The experiments on coherence with 1~GiB data slots and on 
query processing use the block design of Figure~\ref{fig:overview}(b) (Appendix~\ref{app:hardware}),
in which the host is connected through PCIe Gen3 x8, and the program slots, twelve 1~GiB data slots and the shared buffer 
reside in the PL DDR4 behind MIG\_0. 
The micro-benchmarks of Sections~\ref{sec:eval:overhead} and~\ref{sec:eval:coherence}, were measured 
with a version of \delilah, in which four 128~MiB data slots resided in the PS DDR. 
The memory latencies of Figure~\ref{fig:memlat} were measured from the device processors, 
on the PS DDR and on the PL DDR that holds the slots. 

We run experiments on the Daisy platform equipped with a single SSD (Samsung 970 EVO),
connected to a host via PCIe Gen3 x8.
The host is equipped with an Intel Core i7-4790 CPU at 3.60~GHz with 4 cores and 32~GB DDR3, and runs Ubuntu 22.04.
For the query processing experiments, the host runs Linux 6.2.6, and clang 14.0 compiles the eBPF programs;
the micro-benchmarks were run with Linux 5.19.17 and liburing 2.2.
A development server (Intel Core i7-12700KF with 12 cores and 32~GB DDR4, Ubuntu 16.04, Linux 4.15.0)
is used to synthesize the block design and to load the FPGA bitstream and OS image onto Daisy before experiments.
All query processing experiments were repeated ten times, and we report averages.

For query processing, we use the Star Schema Benchmark (SSB). 
The columns of the base tables are stored on the SSD attached to Daisy and processed as described 
in Section~\ref{sec:qp}. We use a scale factor of 10 and one file per column.

\subsection{Cost of Invoking an Offloaded Function}
\label{sec:eval:overhead}

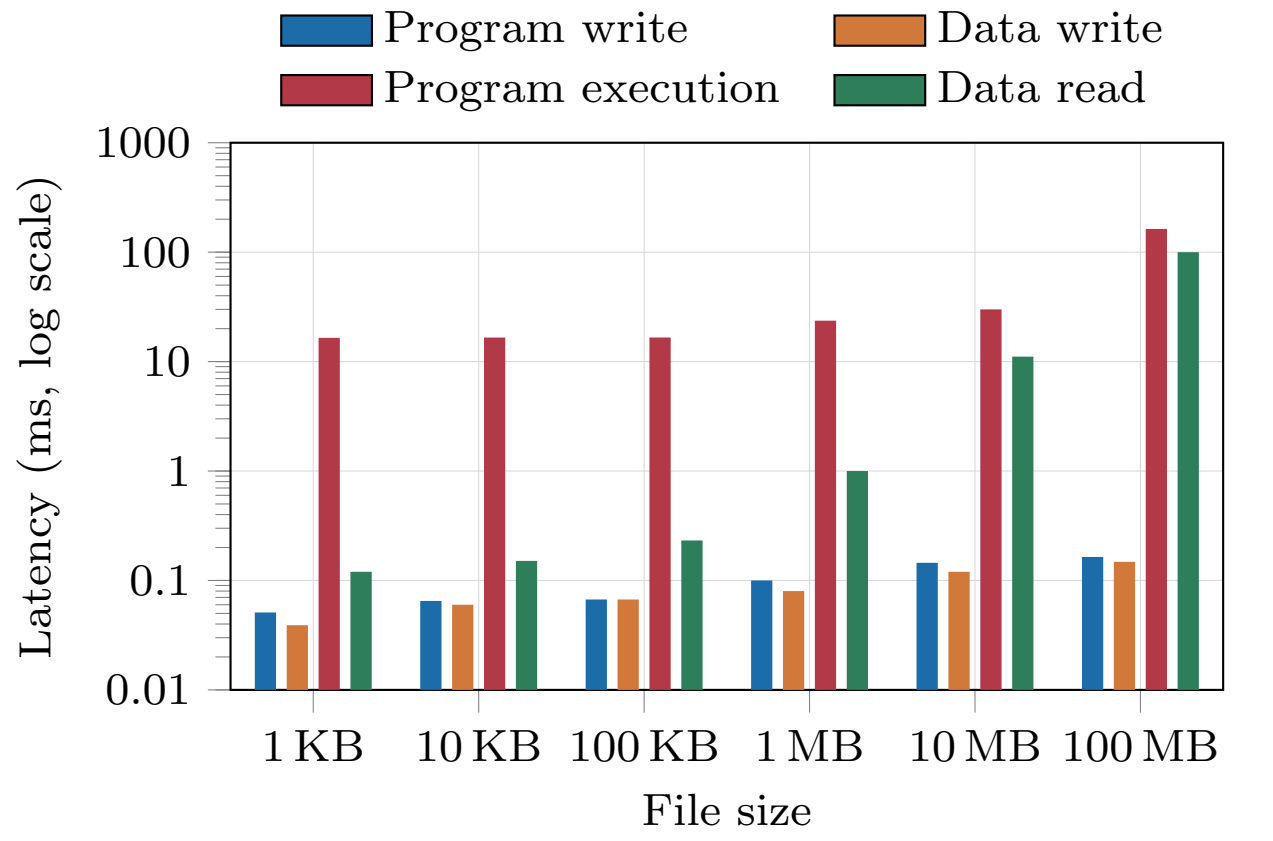
\begin{figure}[htbp]
    \begin{tikzpicture}
\begin{axis}[
  delilah base, ymode=log, delilah log ms,
  width=0.97\linewidth, height=5.2cm,
  ybar, bar width=4pt,
  log origin=infty,
  ymin=0.01, ymax=1000,
  ytick={0.01,0.1,1,10,100,1000},
  symbolic x coords={0.001,0.01,0.1,1,10,100},
  xtick=data,
  xticklabels={1\,KB,10\,KB,100\,KB,1\,MB,10\,MB,100\,MB},
  xlabel={File size},
  enlarge x limits=0.1,
  area legend,
  legend columns=2,
  legend style={at={(0.5,1.02)}, anchor=south, /tikz/every even column/.append style={column sep=8pt}},
]
\addplot[fill=cProg, draw=none] table[x=size_mb, y=prog_write] {latency.dat};
\addlegendentry{Program write}
\addplot[fill=cMeta, draw=none] table[x=size_mb, y=data_write] {latency.dat};
\addlegendentry{Data write}
\addplot[fill=cExec, draw=none] table[x=size_mb, y=prog_exec] {latency.dat};
\addlegendentry{Program execution}
\addplot[fill=cRead, draw=none] table[x=size_mb, y=data_read] {latency.dat};
\addlegendentry{Data read}
\end{axis}
\end{tikzpicture}
    \caption{End-to-end latency (in ms) of the four \delilah driver commands, with whole-slot maintenance 
    (128~MiB data slots in PS DDR).}
    \label{fig:offload}
\end{figure}

To measure the cost of invoking an offloaded function, we use an eBPF program that takes a file name as argument 
and calls a registered function that reads this file into the data slot. 
The size of the file varies from 1~KB to 100~MB. 
Figure~\ref{fig:offload} shows the end-to-end latency of the four \delilah driver commands 
(Section~\ref{sec:delilah:protocol}): writing the eBPF program to a program slot, writing its arguments to a data slot, 
executing the program, and reading the result.

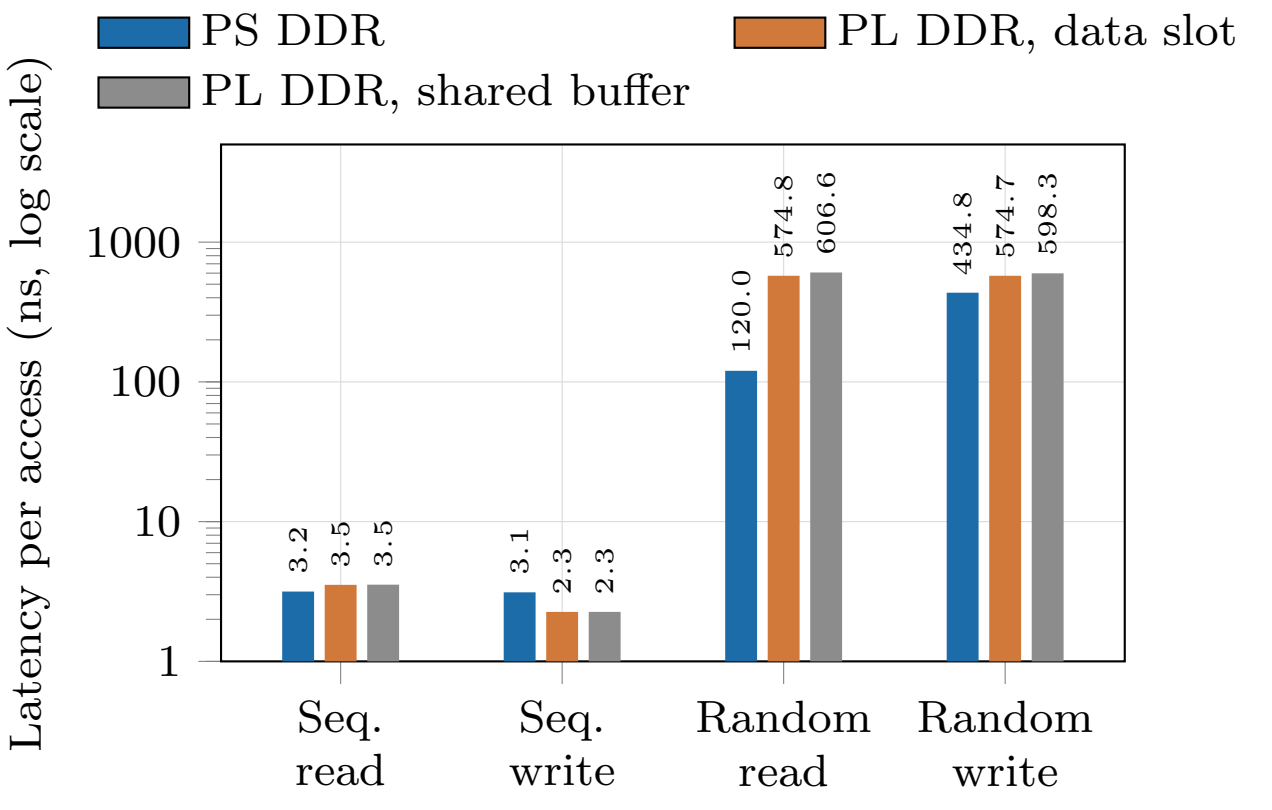
\begin{figure}[htbp]
    \begin{tikzpicture}
\begin{axis}[
  delilah base, ymode=log,
  log ticks with fixed point, /pgf/number format/1000 sep={},
  width=0.9\linewidth, height=5cm,
  ybar, bar width=6pt,
  log origin=infty,
  ymin=1, ymax=5000,
  ytick={1,10,100,1000},
  ylabel={Latency per access (ns, log scale)},
  symbolic x coords={seq_read,seq_write,rand_read,rand_write},
  xtick=data,
  xticklabels={Seq.\\read, Seq.\\write, Random\\read, Random\\write},
  xticklabel style={align=center},
  enlarge x limits=0.18,
  nodes near coords, point meta=rawy,
  nodes near coords style={font=\tiny, rotate=90, anchor=west, /pgf/number format/.cd, fixed, precision=1, zerofill},
  area legend,
  legend columns=2,
  legend style={at={(0.5,1.02)}, anchor=south, /tikz/every even column/.append style={column sep=6pt}},
]
\addplot[fill=cProg, draw=none] table[x=access, y=ps_local] {memory-latency.dat};
\addlegendentry{PS DDR}
\addplot[fill=cMeta, draw=none] table[x=access, y=pl_dslot] {memory-latency.dat};
\addlegendentry{PL DDR, data slot}
\addplot[fill=cNeutral, draw=none] table[x=access, y=pl_sslot] {memory-latency.dat};
\addlegendentry{PL DDR, shared buffer}
\end{axis}
\end{tikzpicture}
    \caption{Average latency (in ns, log scale) of one access from the device processors to the PS DDR and to 
    the PL DDR that holds the data slots and the shared buffer, for sequential and random reads and writes 
    (caches enabled, 32~M accesses).}
    \label{fig:memlat}
\end{figure}

Writing the program and its arguments is fast, and reading the result takes time proportional to its size, 
as expected for DMA transfers. 
Program execution, however, has a high floor of about 16.5~ms, regardless of the size of the file.
Program execution differs from the other commands in that it involves a command sent through BAR0 and the execution 
of the program in uBPF. 

For the program that reads a 1~KB file, the latency of execution breaks down as follows. 
The host driver, the execution of the eBPF program and of the registered function, and interrupt handling all 
contribute little (0.54~ms together). Most of the time is spent invalidating (8.47~ms) and flushing (7.93~ms)
the cache of the device processors for the DMA buffers that hold the program and data slots,
97\% of the 16.94~ms end-to-end latency. Without further information, the controller maintains coherence over 
the entire slots, so this time is proportional to the size of the slots (128~MiB data slots in this version 
of \delilah), not to the amount of data. We refer to this as {\em whole-slot maintenance}.

\begin{figure}[htbp]
    \begin{tikzpicture}
\begin{axis}[
  delilah base, ymode=log, delilah log ms,
  width=0.97\linewidth, height=5.2cm,
  xmode=log, log basis x=2,
  xmin=24, xmax=5.5e6,
  ymin=0.0007, ymax=1000,
  ytick={0.001,0.01,0.1,1,10,100,1000},
  xtick={32,1024,32768,1048576},
  xticklabels={32\,B,1\,KB,32\,KB,1\,MB},
  xlabel={Transfer size},
  unbounded coords=discard,
  legend columns=2,
  legend style={at={(0.5,1.02)}, anchor=south, /tikz/every even column/.append style={column sep=6pt}},
]
\addplot[delilah line, color=cProg, mark=*] table[x=bytes, y=dma_w] {cmb.dat};
\addlegendentry{DMA write}
\addplot[delilah line, color=cProg, dashed, mark=o, mark options={solid}] table[x=bytes, y=dma_r] {cmb.dat};
\addlegendentry{DMA read}
\addplot[delilah line, color=cMeta, mark=square*] table[x=bytes, y=cmb_w] {cmb.dat};
\addlegendentry{CMB write}
\addplot[delilah line, color=cMeta, dashed, mark=square, mark options={solid}] table[x=bytes, y=cmb_r] {cmb.dat};
\addlegendentry{CMB read}
\end{axis}
\end{tikzpicture}
    \caption{Latency of transfers between host and device through DMA and through the memory-mapped BAR0 region (CMB).
    CMB writes of 32~B and 512~B took less than the 1~\textmu s timer resolution and are not shown.}
    \label{fig:memory}
\end{figure}
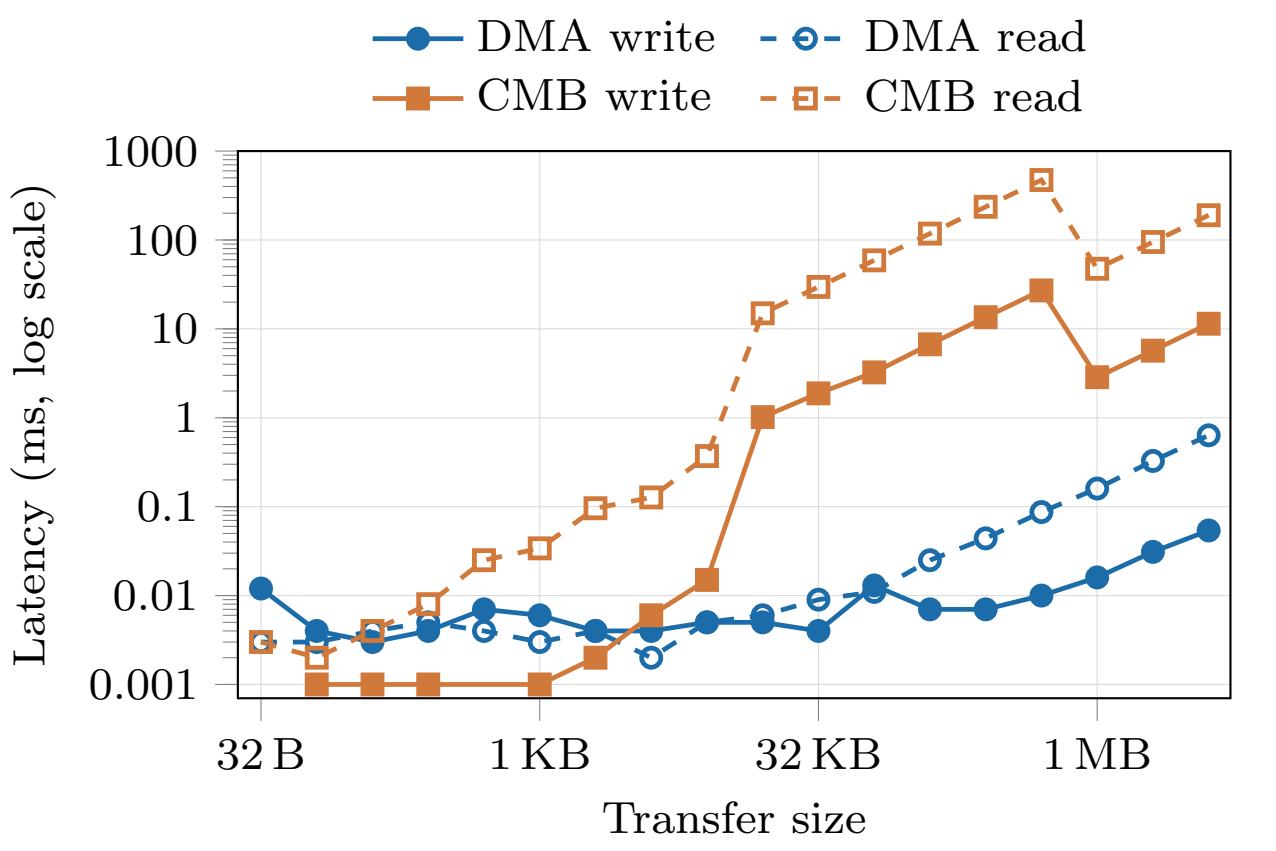
The program used in these micro-benchmarks is deliberately minimal: it calls the registered function 
that reads a file into the data slot and never touches the data itself. 
The programs used for query processing (Section~\ref{sec:qp}, Figures~\ref{lst:filter}--\ref{lst:convert}) 
are different: they scan their input, write intermediate results (bitmasks, converted years) back to device memory 
and, for aggregation, gather values at the positions selected by the filters. 
Their cost is thus dominated by memory accesses from the device processors, 
and on the Zynq UltraScale+ these are cheap only when they are sequential (Figure~\ref{fig:memlat}). 
With caches enabled, a sequential access takes 2--4~ns, whereas a random access takes 575--607~ns to the PL DDR 
that holds the program and data slots and the shared buffer in our block design (Figure~\ref{fig:overview}) 
(and 120--435~ns to the PS DDR), two orders of magnitude more. 
This is consistent with prior measurements showing that the memory performance of Zynq UltraScale+ systems 
varies widely, and often unexpectedly, with the access pattern and the path to memory~\cite{manev_unexpected_2019}. 
We return to the cost of the query operators in Section~\ref{sec:eval:qp}.

\subsection{The Coherence Trade-off}
\label{sec:eval:coherence}
\label{sec:eval:transfers}

Section~\ref{sec:eval:overhead} shows that, with whole-slot maintenance, the cost of invoking an offloaded function 
is the cost of keeping program and data slots coherent, regardless of how much data the program uses. 
This cost grows with the size of the slots and with the distance to memory: with the 1~GiB data slots in the PL DDR4, 
a single maintenance pass over the data slot takes about 120~ms, i.e., about 240~ms per GiB, against 130~ms per GiB 
in the PS DDR, consistent with the slower path to the PL DDR (Figure~\ref{fig:memlat}). 
The options of Section~\ref{sec:delilah:coherence} move this cost elsewhere (question~(b)). 
Two simple alternatives fail. Non-cacheable slots impose alignment constraints that the ELF files of eBPF 
programs do not respect. Copying data between separate DMA buffers and the slots beats whole-slot maintenance 
for small transfers (0.21~ms instead of 16.5~ms for 1~KB), but its cost grows superlinearly beyond 10~MB 
(1.5~s instead of 163~ms for 100~MB). 
Range-based maintenance makes the cost of coherence proportional to the data exchanged: with 1~GiB data slots, 
executing the program that reads a 1~KB file takes 0.27~ms instead of about 240~ms, but the host must declare, 
at cache-line granularity, the ranges it exchanges. 
Hardware coherence would remove both the maintenance and the declarations, at the price of confining the slots 
to the small PS DDR and routing all transfers through the coherent ports.

The one coherent path that \delilah already uses is BAR0, which resides in the PS DDR and is reached by the host 
through S\_AXI\_HPC0\_FPD. 
Besides DMA, the host can thus access device memory through loads and stores to this memory-mapped region, 
the mechanism used to access the controller memory buffer (CMB) of NVMe devices (question~(c)). 
Figure~\ref{fig:memory} compares the two mechanisms for reads and writes of 32~B to 4~MB. 
CMB writes take at most 2~\textmu s up to 2~KB, faster than DMA, while DMA writes are faster from 4~KB on; 
CMB reads are on par with DMA reads only up to 128~B and slower beyond. 
CMB accesses need no cache maintenance, but they reach a small region of the PS DDR rather than the slots, 
and they are only efficient for small updates, e.g., log records~\cite{lee_x-ssd_2022}. 
This is the trade-off in miniature: coherent memory is cheap to keep coherent but scarce, 
whereas the large slots in the PL DDR4 must be kept coherent by the host and the controller.

\subsection{Query Processing}
\label{sec:eval:qp}

We now turn to question~(d) and offload SSB query Q3.3 to \delilah, at scale factor~10. 
Q3.3 joins the \texttt{lineorder} fact table with the \texttt{customer}, \texttt{supplier} and 
\texttt{date} dimensions, selects customers and suppliers from two given cities over a range of years, 
and aggregates revenue grouped by customer city, supplier city and year.

\paragraph{Methodology.}
We hand-code the offload of Q3.3 rather than generating it from a storage manager.
On this query, we run a series of experiments~\cite{hedam_delilah_thesis_2024}:
full offload with Neon filters (see Section~\ref{sec:qp}), full offload with the FPGA filter accelerator, city filters run concurrently
and sequentially, and host-only execution as a baseline.
Each experiment isolates one factor, so that the breakdown below can attribute time to staging, operators,
transfers and offload mechanisms.

\paragraph{Data layout.}
Each column used by the query is stored as a separate file on the SSD attached to Daisy, with one value per 
\texttt{lineorder} row. The customer and supplier cities are denormalized into the fact table and dictionary-encoded 
as 32-bit integers; the dictionary is kept on the host, which translates the two city names into codes. 
The order date is stored as 8~characters (YYYYMMDD), and the revenue as a 32-bit integer. 
At scale factor~10, each column holds 60~million values, i.e., 240~MB (480~MB for the order date).

\paragraph{Execution.}
We offload all the operators that are implemented as eBPF programs (full offload): 
the four city predicates (two per city column), the conversion of the order date into a year, 
the filter on the year, and the aggregation of revenue. 
Filters produce bitmasks, and use Neon SIMD instructions. The four city predicates run concurrently on four 
uBPF virtual machines, and all eBPF programs are JIT compiled. 

The joins are executed on the host. Because the dimension attributes are denormalized, a join reduces to combining 
the bitmasks returned by the filters position by position; the host then groups the qualifying positions by 
customer city, supplier city and year, invokes the aggregation on the device once per group, and projects the result. 
The intermediate results that travel to the host are the bitmasks of the filters, and the converted years used for grouping. 
At scale factor~10, the four columns used by the query fit in the 1~GB shared buffer: each is staged once, 
before the first operator, and later passes read the staged copy.

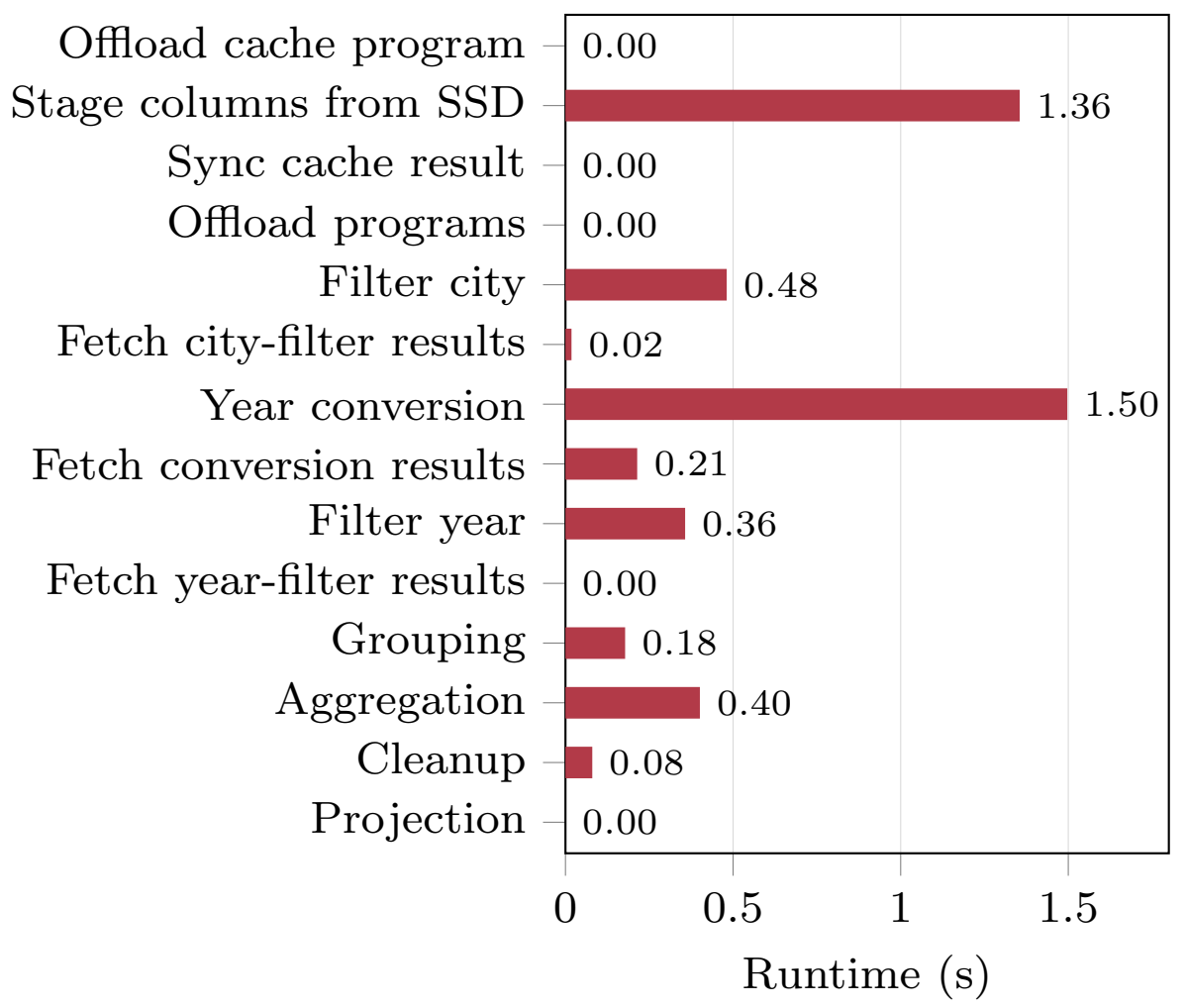
\begin{figure}[htbp]
    \begin{tikzpicture}
\begin{axis}[
  delilah base,
  scale only axis, width=0.48\linewidth, height=5.6cm,
  xbar, bar width=6pt,
  xmin=0, xmax=1.8,
  xlabel={Runtime (s)},
  symbolic y coords={Offload_cache,Cache_execution,Sync_Cache_result,Offload_programs,
    Filter_city_execution,Fetch_filtering_results,Conversion_execution,Fetch_conversion_results,
    Filter_year_execution,Fetch_year_filtering_results,Grouping,Aggregation,Cleanup,Projection},
  ytick=data,
  yticklabels={Offload cache program, Stage columns from SSD, Sync cache result, Offload programs,
    Filter city, Fetch city-filter results, Year conversion, Fetch conversion results,
    Filter year, Fetch year-filter results, Grouping, Aggregation, Cleanup, Projection},
  y dir=reverse,
  enlarge y limits=0.04,
  ymajorgrids=false,
  nodes near coords,
  nodes near coords style={font=\scriptsize, black, /pgf/number format/.cd, fixed, precision=2, zerofill},
]
\addplot[fill=cExec, draw=none] table[x=s, y=stage] {ssb-q33-breakdown.dat};
\end{axis}
\end{tikzpicture}
    \caption{Breakdown of the runtime (in s) of SSB Q3.3 (SF~10) with all eBPF programs offloaded to \delilah, 
    with Neon SIMD filtering and the shared buffer in PL memory, in execution order. Joins, grouping and projection 
    are executed on the host. Data from~\cite{hedam_delilah_thesis_2024}, read off the published figure 
    with a resolution of 0.01~s per stage.}
    \label{fig:ssb}
\end{figure}

Figure~\ref{fig:ssb} breaks down the runtime of the query, about 4.58~s in total,
as measured in the experimental campaign of~\cite{hedam_delilah_thesis_2024} (averages over ten runs).
Memory accesses dominate cost: staging the four columns from the SSD into the shared buffer accounts for 1.36~s, 
30\% of the runtime, while the conversion operator, which reads the order dates from the shared buffer and writes 
the years to the data slot, accounts for 1.50~s, 33\%. 
The filters take 0.84~s together (18\%) with Neon vectorization, and aggregation 0.40~s (9\%). 
The host-side stages, grouping (0.18~s) and cleanup (0.08~s), and the transfers of intermediate results to the host 
(0.23~s, 5\%) account for the rest. 
The costs that are specific to eBPF offload (offloading programs, offloading and synchronizing the cache) 
are below 0.01~s.

\paragraph{Hardware-accelerated filtering.}
We evaluated the filter accelerator of Section~\ref{sec:delilah:hwfilter}. Its gains are modest. This experiment compares
the accelerator with eBPF filters that do not use Neon. On the full SSB Q3.3 query, with the city
filters running concurrently, hardware filtering reduces the runtime from 5.85~s to 5.72~s; with the filters run one
after the other, from 7.29~s to 6.46~s. The accelerator is fast, but it shares the path to memory with the other 
operators, which saturates first. Developing it confirmed this: almost all of its time was spent waiting for memory, 
bundling accesses into vector copies brought the main speedup, whereas loop unrolling brought none and costs 
FPGA real estate, since unrolled iterations execute concurrently. Each accelerator also adds LUTs, flip-flops and 
routing pressure to a design that is already hard to place (Section~\ref{sec:delilah:synthesis}), and debugging is 
difficult: an automatically inferred interface setting that forced 64-bit aligned accesses on 32-bit data was 
detected by neither the HLS tools nor simulation. As with the software filters, the bottleneck is the data path to 
memory, not computation.

\paragraph{Host baseline.}
We compare against a baseline in which the host executes the whole query~\cite{hedam_delilah_thesis_2024}. 
Each column is read once from the SSD through \delilah (an eBPF program calls the registered function that reads 
the file into device memory, and the host fetches it), and all operators run on the host CPU. 
This baseline takes about 5.55~s. In the same experiment, full offload takes about 5.0~s, so offloading makes the
query about 10\% faster; with the best filter implementation (Figure~\ref{fig:ssb}, 4.58~s), about 17\% faster.
The gain does not come from processing: the host executes all its operators in about 1.5~s, whereas filtering, 
conversion and aggregation take about 2.7~s on the device. It comes from data movement: the host-only execution 
spends about 4.0~s moving the columns from the SSD through device memory to host memory, paying for coherence 
on each transfer, whereas offloaded operators keep the data in device memory and only ship intermediate results.

\paragraph{Chunked execution.}
The run of Figure~\ref{fig:ssb} is a best case: at scale factor~10 the columns fit in the shared buffer, so each
operator runs to completion over a whole column in a single invocation. Larger data sets require running each
operator over chunks, as described in Section~\ref{sec:qp}. The experimental campaign of~\cite{hedam_delilah_thesis_2024}
measures this directly, by running the full offload of Q3.3 either to completion or in chunks of 128~MB.
Chunking increases the runtime from about 5.0~s to about 8.1~s, i.e., by about 60\%, which makes offload slower
than host-only execution (5.55~s).
First, every chunk requires a round of host-device commands: cache maintenance and JIT compilation are paid per chunk
rather than per operator, and the device is idle while the host decides what to do next.
Second, staging is no longer a separate stage but is folded into the operators that read each chunk from the SSD:
the city filters take about 1.7~s and the conversion about 3.1~s, instead of 1.5~s.
Third, aggregation takes about 2.1~s instead of 0.4~s, because its input arrives in chunks and it makes several
passes over the same data.
In addition, the shared buffer must be allocated and managed chunk by chunk, without knowing the size of the columns
in advance. Chunked execution thus multiplies the costs that the model imposes: host-device round trips, coherence
maintenance and passes over device memory.

The micro-benchmarks thus carry over: with range-based maintenance, invoking an eBPF program and transferring its results cost little. 
They do not, however, capture the cost of the operators themselves, which, unlike the program used 
in the micro-benchmarks, access device memory heavily (Section~\ref{sec:eval:overhead}, Figure~\ref{fig:memlat}).
The problem lies in the execution model.
Because the host drives the loop across chunks and the sequence of operators, each operator is a separate pass over 
its input: in general, chunks are staged from the SSD for each pass, intermediate results are materialized in device memory 
or shipped to the host, and the state carried from one chunk to the next is passed as arguments through the data slot. 
Nothing is pipelined across operators on the device. 
The time is spent moving and reorganizing data within device memory, the scarce resource, rather than processing it.

\subsection{Discussion}
We do not claim that computational storage never pays off, but that the standard execution model does not make it pay
off for query processing, and that a single query is enough to show it.
First, the roadblocks are structural. Host-driven iteration over chunks, state carried through the data slot,
operators invoked one pass at a time and intermediate results materialized between them are imposed by the model,
so every query inherits them.
Second, Q3.3 is a favorable case. It is selective, it reads few columns, its operators are simple, and at scale
factor~10 its columns fit in the shared buffer, so that each is staged only once.
Yet offloading saves less than a fifth of the runtime of host execution, and only because it avoids data transfers
to the host; the operators themselves run slower on the device. Neither faster software filters nor hardware
acceleration changes this, because the bottleneck is the path to device memory.
Third, integration would make things worse, not better. Larger scale factors require running each operator over chunks,
which already makes Q3.3 about 60\% slower than running each operator to completion, and slower than host-only
execution (Section~\ref{sec:eval:qp}); wider plans would multiply host-device round trips.
Other queries would change the relative weight of the operators, but not the structure of the execution,
which is set by the model.
We therefore stopped before integrating \delilah into a storage manager and running more complete workloads:
further engineering in this direction would measure the same roadblocks at a larger scale.

\section{Perspectives}
\label{sec:perspectives}

\delilah implements computational storage as exposed by the standard interfaces, with one deviation: 
registered functions, rather than host-issued storage requests, load data from the SSDs into device memory. 
This deviation takes the host out of the movement of base data, so a device that follows the standards strictly 
would incur at least the costs we measured. The failure we observe stems from host-driven iteration, 
which \delilah shares with the standard model. 
Our experience with it leads to lessons at three levels: the representation of offloaded code, the execution model, 
and the architecture of the device itself.

\paragraph{eBPF is glue, not the program.}
In \delilah, eBPF plays the role of a relocatable object file. An eBPF program is linked on the device against registered 
functions. This way, the host can call any registered 
function without extending the host-device interface, and the cost of the eBPF program itself is small. 
The registered functions, however, which do the actual work, are not implemented in eBPF. They are native object files, 
compiled for the ARM cores or bound to accelerators on the FPGA, and linked into the controller when the device is 
configured. A query offloaded to \delilah is therefore expressed in two representations: downloadable eBPF glue, 
and device-defined functions in native code, with different toolchains, life cycles and safety properties.

\paragraph{Memory belongs in the protocol.}
What decides performance is memory. On a device whose processors access shared buffers through non-coherent caches, 
minimizing copies and keeping DMA buffers coherent pull in opposite directions. Declaring the ranges exchanged 
with the device makes the cost of coherence proportional to the data exchanged, but makes how device memory is used 
part of the host-device protocol; hardware coherence hides it from the protocol, but confines shared buffers to 
the small memory of the device processors.

\paragraph{The host should not drive the iteration.}
Our main lesson concerns the execution model:
Computational storage based on host-invoked functions and host-managed 
data movement is not a good match for query processing.
When the host manages the loop across chunks, carries state from one chunk to the next, and invokes operators one 
after the other, execution degenerates into column-at-a-time processing in chunks, and the time goes into staging and 
reorganizing data in device memory. Batch requests would reduce the number of interactions but not this structure. 
The device must manage the iteration over data and its movement, as anticipated by SNIA's indirect usage and 
variable-sequence aggregator functions, and as demonstrated by programmable SSDs over proprietary 
interfaces~\cite{gu_biscuit_2016}.

\paragraph{A general-purpose software stack is hard to scale.}
Moving the iteration onto \delilah would not be enough, because its software stack would then sit on the critical path 
of every I/O. Even when the host declares the ranges it exchanges, invoking an offloaded program costs on the order of 
a quarter of a millisecond, spent in Linux, in the controller, in uBPF and in interrupt handling on the device. 
These costs are specific to our platform, with Cortex-A53 cores, a general-purpose operating system, a virtual machine 
in user space and interrupt-driven completion; faster cores, polling, batching or kernel bypass would reduce them, 
perhaps by an order of magnitude.

\paragraph{The program representation does not fit the device.}
eBPF was designed for short, stateless functions that run to completion on one event. 
The programs that query processing offloads are different: they are long-running, they keep state across chunks 
(selection vectors, hash tables, partial aggregates), and they must exploit all the cores of the device concurrently. 
\delilah bridges this gap outside eBPF. Each uBPF virtual machine runs in a Linux thread pinned to a core, 
concurrency is obtained by running independent programs on different virtual machines, and state is carried across 
invocations through the data slot. The representation of offloaded code thus captures neither the concurrency nor 
the state of the computation, which are left to the operating system and to the host. 
Going forward, two changes appear necessary. First, as mentioned above,
downloadable and registered functions should share a unified 
representation, so that the same toolchain, linking model and verification apply to the glue and to the operators, 
and so that the boundary between device-defined and downloadable code can move as accelerators evolve. 
Second, this representation should map efficiently onto the cores and accelerators of the device: it should express 
concurrent, stateful programs, with explicit state that persists across events and explicit communication between 
computation sites, rather than isolated functions scheduled by a general-purpose operating system.

\paragraph{A single computation site misses the flow.}
The standards model a computational storage device as a computation site: an engine, with its execution environment, 
executes a function on data staged in function data memory. \delilah faithfully reproduces this store-then-process 
structure. Data in an SSD, however, is not a buffer but a flow~\cite{Lerner27}. It moves in packets, from flash channels through 
the controller and its memory to the host interface, and each stage of this path offers a natural place to compute: 
per-channel pattern matching, record parsing and filtering in the controller, 
aggregation and re-encoding before the host interface. 
Like packet processing on smart NICs, near-data processing on SSDs calls for multiple 
computation sites along the data path, each with a small budget per packet, rather than a single general-purpose 
engine at the end of a staging step. Staging data in a single memory to compute on it is precisely what makes memory 
the scarce resource.

\paragraph{Towards co-designed SSDs.}
These lessons lead us from computational storage, which accesses SSDs through their external interface and has no 
advantage over the host when accessing stored data~\cite{lerner_not_2021}, to SSDs co-designed with the data 
systems they serve~\cite{Lerner2024-kb}. 
In the CHORYS project, we pursue this direction with OpenSSD-V~\cite{bonnet_chorys_2026}, an open computational SSD in 
which RISC-V cores coupled with data-pattern accelerators replace the fixed-function controller. 
Its platform places several compute sites along the path between the host interface and the flash interface, and an 
NVMe command can be guided through an accelerated pipeline that, e.g., parses a Parquet stream, filters the records and 
re-encodes the survivors, so that the I/O returns the result of a pushed-down operator rather than raw blocks. 
The iteration is driven by the data path rather than by the host. The control path is designed for high event rates, 
with hardware sequencers that configure accelerators without per-task register writes and lightweight signalling 
that avoids trapping into a kernel. On the host side, operators exchange columnar batches according to Substrait plans, 
so that a device can return filtered batches, bitmaps or partial aggregates without changing the plan.

Some lessons of \delilah carry over directly. A relocatable format remains the right way to deliver code to the device, 
but it should cover both downloadable and device-defined functions, and describe concurrent, stateful programs that 
configure the compute sites, rather than functions interpreted or dispatched per event. 
Memory, and its coherence, must be designed into the device and its interface rather than left to the host. 
And safety, which \delilah left out of scope (Section~\ref{sec:delilah:design}), becomes more pressing when the SSD is programmable.

\section{Conclusion}
\label{sec:conclusion}

Computational storage is now standardized. The SNIA programming model is broad, but the standard interfaces that 
realize it, the SNIA Computational Storage API and the NVMe Computational Programs and Subsystem Local Memory 
command sets, expose a narrower execution model: the host invokes functions on the device, on data that the host 
has moved into device memory. Both standards single out eBPF to represent the functions that the host downloads 
at run-time. In this paper, we asked whether this form of computational storage is a good match for query processing.

To answer this question, we designed and implemented \delilah, a computational storage system that follows the 
host-orchestrated execution model on an actual device, the OpenSSD Daisy platform, and runs eBPF programs that call 
registered functions on the device.
We hand-coded the offload of a single SSB query as a first step toward integrating \delilah into a storage manager,
and ran a series of experiments on it. That step was sufficient.
Our evaluation leads to three conclusions.
First, eBPF is best understood as a relocatable binary format: it is a lightweight mechanism for calling registered, 
possibly hardware-accelerated, functions without extending the host-device interface, and its overhead is negligible. 
Second, the mechanisms of the execution model are dominated by memory, not compute. The coherence of the buffers shared 
between host and device dominates the latency of offloaded functions. Keeping them coherent is a trade-off: 
software maintenance costs time proportional to the size of the shared buffers, unless the host declares the ranges 
it exchanges, which makes coherence part of the host-device protocol; hardware coherence confines the shared buffers 
to the small memory of the device processors. 
Third, host-invoked functions and host-managed data movement are not a good match for query processing. 
When the host drives the iteration over chunks and the sequence of operators, execution degenerates into 
column-at-a-time processing in chunks, and the time goes into staging and reorganizing data in device memory 
rather than into processing it. On SSB Q3.3, offloading saves less than a fifth of the runtime of host execution, 
and only because it avoids moving the data to the host: the operators themselves run slower on the device.
Running the operators over chunks, as larger data sets require, makes the offloaded query about 60\% slower,
and slower than host execution.
These roadblocks are structural. They appear on a query that is favorable to offload and with data that fits
in device memory, and storage-manager integration would only amplify them.
They are a show-stopper for host-orchestrated computational storage in query processing,
which is why we did not pursue this direction further.

Query processing on computational storage thus requires an execution model in which the device manages the iteration 
over data and its movement. The SNIA programming model anticipates such a model with indirect usage and variable-sequence 
aggregator functions, but the standard interfaces do not expose it yet. 
More importantly, a single general-purpose computation site, running an operating system and an eBPF virtual machine, 
is hard-pressed to sustain the millions of events per second of modern flash and PCIe, and it ignores the flow nature of data on 
its way from flash to host. We therefore argue for SSDs co-designed with data systems, with multiple computation sites 
along the data path, which is the direction we now pursue with OpenSSD-V.

\bibliographystyle{abbrv}
\bibliography{main}
\appendix
\appendix

\section{\delilah Implementation Details}
\label{app:overview}
\label{app:components}

This appendix details the implementation of \delilah: its software components
on host and Zynq PS, the hardware components of the PL,
and the layout of device memory.
The code of all components is available in the \delilah project on GitHub\footnote{\url{https://github.com/delilah-csp}};
we give a pointer to the repository of each component below.
This appendix, generated with Claude Code from the code base, is essentially a guide to the Delilah github repository.
The scripts and programs used for the experimental evaluation are available in a separate
repository\footnote{\url{https://github.com/delilah-csp/delilah-eval}}.

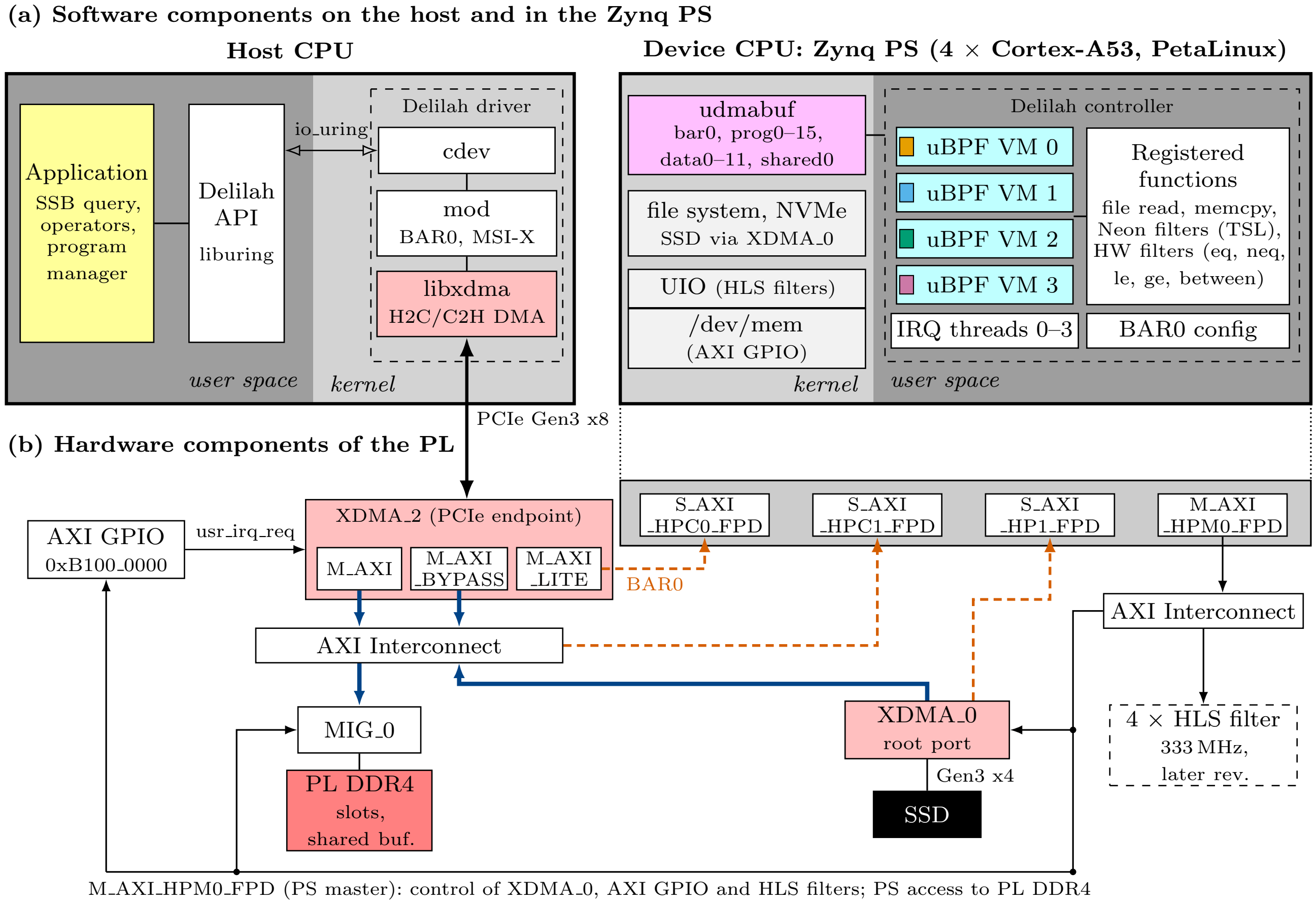
\begin{figure*}[tbp]
  \centering
  \definecolor{aUser}{HTML}{9E9E9E}
\definecolor{aKern}{HTML}{D3D3D3}
\definecolor{aApp}{HTML}{FFFF99}
\definecolor{aXdma}{HTML}{FFBFBF}
\definecolor{aUdma}{HTML}{FFBFFF}
\definecolor{aUbpf}{HTML}{BFFFFF}
\definecolor{aLnx}{HTML}{F2F2F2}
\definecolor{bdBlue}{HTML}{004488}
\definecolor{bdVerm}{HTML}{D55E00}
\definecolor{bdRam}{HTML}{FF8080}
\definecolor{bdPS}{HTML}{CCCCCC}
\definecolor{eZero}{HTML}{E69F00}
\definecolor{eOne}{HTML}{56B4E9}
\definecolor{eTwo}{HTML}{009E73}
\definecolor{eThree}{HTML}{CC79A7}
\begin{tikzpicture}[
  font=\footnotesize,
  box/.style={draw=black, line width=0.5pt, align=center, inner sep=2pt},
  app/.style={box, fill=aApp},
  dl/.style={box, fill=white},
  lnx/.style={box, fill=aLnx},
  xd/.style={box, fill=aXdma},
  ud/.style={box, fill=aUdma},
  vm/.style={box, fill=aUbpf},
  port/.style={box, fill=white, font=\scriptsize, inner sep=1pt},
  link/.style={draw=black, line width=0.6pt},
  zone/.style={font=\footnotesize\itshape},
  ptitle/.style={font=\footnotesize\bfseries, anchor=west, inner sep=0pt},
  title/.style={font=\footnotesize\bfseries},
  sub/.style={font=\scriptsize},
  blue/.style={draw=bdBlue, line width=1.6pt, -{Latex[length=2mm]}},
  verm/.style={draw=bdVerm, line width=1pt, dash pattern=on 3pt off 2pt, -{Latex[length=2mm]}},
  ctl/.style={draw=black, line width=0.5pt, -{Latex[length=1.6mm]}},
  row/.style={draw=black, line width=0.4pt},
  psm/.style={draw=black, line width=0.6pt, -{Latex[length=1.6mm]}},
]

\node[ptitle] at (0,0.75) {(a) Software components on the host and in the Zynq PS};

\node[title] at (3.7,0.3) {Host CPU};
\fill[aUser] (0,0) rectangle (4.0,-4.3);
\fill[aKern] (4.0,0) rectangle (7.4,-4.3);
\draw[line width=1.2pt] (0,0) rectangle (7.4,-4.3);
\node[zone, anchor=south east] at (3.9,-4.25) {user space};
\node[zone, anchor=south west] at (4.1,-4.25) {kernel};

\node[app, text width=1.6cm, minimum height=3.1cm] (app) at (1.05,-1.95)
  {Application\\[3pt]{\scriptsize SSB query,\\operators,\\program\\manager}};
\node[dl, text width=1.1cm, minimum height=3.1cm] (api) at (3.0,-1.95)
  {Delilah\\API\\[3pt]{\scriptsize liburing}};
\draw[link] (app) -- (api);

\draw[black, line width=0.5pt, dashed] (4.75,-0.2) rectangle (7.25,-3.75);
\node[sub, anchor=north] at (6.0,-0.22) {Delilah driver};
\node[dl, minimum width=2.3cm, minimum height=0.6cm] (cdev) at (6.0,-1.0) {cdev};
\node[dl, text width=2.2cm, minimum height=0.85cm] (mod) at (6.0,-1.95)
  {mod\\{\scriptsize BAR0, MSI-X}};
\node[xd, text width=2.2cm, minimum height=0.85cm] (hx) at (6.0,-3.0)
  {libxdma\\{\scriptsize H2C/C2H DMA}};
\draw[link] (cdev) -- (mod);
\draw[link] (mod) -- (hx);
\draw[draw=black, fill=white, line width=0.5pt, {Latex[length=2mm, open]}-{Latex[length=2mm, open]}]
  (api.east |- cdev) -- node[above=1pt, sub] {io\_uring} (cdev.west);

\node[title] at (12.5,0.3) {Device CPU: Zynq PS (4 $\times$ Cortex-A53, PetaLinux)};
\fill[aKern] (8.0,0) rectangle (11.3,-4.3);
\fill[aUser] (11.3,0) rectangle (17.0,-4.3);
\draw[line width=1.2pt] (8.0,0) rectangle (17.0,-4.3);
\node[zone, anchor=south east] at (11.2,-4.25) {kernel};
\node[zone, anchor=south west] at (11.4,-4.25) {user space};

\node[ud, text width=2.95cm, minimum height=1.0cm] (udm) at (9.65,-0.8)
  {udmabuf\\{\scriptsize bar0, prog0--15,\\data0--11, shared0}};
\node[lnx, text width=2.95cm, minimum height=0.85cm] (nvme) at (9.65,-1.95)
  {file system, NVMe\\{\scriptsize SSD via XDMA\_0}};
\node[lnx, text width=2.95cm, minimum height=0.5cm] (uio) at (9.65,-2.8)
  {UIO {\scriptsize (HLS filters)}};
\node[lnx, text width=2.95cm, minimum height=0.5cm] (dmem) at (9.65,-3.45)
  {/dev/mem {\scriptsize (AXI GPIO)}};

\draw[black, line width=0.5pt, dashed] (11.45,-0.2) rectangle (16.85,-3.75);
\node[sub, anchor=north] at (14.15,-0.22) {Delilah controller};
\foreach \i/\c in {0/eZero,1/eOne,2/eTwo,3/eThree} {
  \node[vm, minimum width=2.3cm, minimum height=0.5cm] (vm\i) at (12.75,-0.95-0.6*\i)
    {\hspace{0.2cm}uBPF VM \i};
  \fill[\c, draw=black, line width=0.4pt] ($(vm\i.west)+(0.05,-0.12)$) rectangle ++(0.18,0.24);
}
\node[dl, minimum width=2.3cm, minimum height=0.45cm] (irq) at (12.75,-3.35) {IRQ threads 0--3};
\node[dl, text width=2.5cm, minimum height=2.3cm, anchor=north] (rf) at (15.4,-0.7)
  {Registered\\functions\\[2pt]{\scriptsize file read, memcpy,\\Neon filters (TSL),\\HW filters (eq, neq,\\le, ge, between)}};
\node[dl, text width=2.5cm, minimum height=0.45cm] (cfg) at (15.4,-3.35) {BAR0 config};
\draw[link] (vm1.east |- rf) -- (rf.west);
\draw[link] (udm.east) -- (11.45,-0.8);

\draw[draw=black, line width=1.2pt, {Latex[length=2.5mm]}-{Latex[length=2.5mm]}]
  (hx.south) -- (6.0,-5.55) node[pos=0.5, right, sub] {PCIe Gen3 x8};

\node[ptitle] at (0,-4.85) {(b) Hardware components of the PL};

\fill[bdPS] (8.0,-5.3) rectangle (17.0,-6.15);
\draw[line width=0.6pt] (8.0,-5.3) rectangle (17.0,-6.15);
\draw[densely dotted, line width=0.6pt] (8.0,-4.3) -- (8.0,-5.3) (17.0,-4.3) -- (17.0,-5.3);
\node[port, minimum width=1.68cm, minimum height=0.55cm] (hpc0) at (9.1,-5.75) {S\_AXI\\\_HPC0\_FPD};
\node[port, minimum width=1.68cm, minimum height=0.55cm] (hpc1) at (11.35,-5.75) {S\_AXI\\\_HPC1\_FPD};
\node[port, minimum width=1.68cm, minimum height=0.55cm] (hp)   at (13.6,-5.75) {S\_AXI\\\_HP1\_FPD};
\node[port, minimum width=1.68cm, minimum height=0.55cm] (hpm0) at (15.85,-5.75) {M\_AXI\\\_HPM0\_FPD};

\draw[line width=0.5pt, fill=aXdma] (3.9,-5.55) rectangle (7.9,-6.85);
\node[sub] at (5.9,-5.78) {XDMA\_2 (PCIe endpoint)};
\node[port, minimum width=1.1cm, minimum height=0.55cm] (maxi)  at (4.6,-6.45)  {M\_AXI};
\node[port, minimum width=1.1cm, minimum height=0.55cm] (mbyp)  at (5.9,-6.45)  {M\_AXI\\\_BYPASS};
\node[port, minimum width=1.1cm, minimum height=0.55cm] (mlite) at (7.2,-6.45)  {M\_AXI\\\_LITE};

\node[dl, text width=1.9cm, minimum height=0.75cm] (gpio) at (1.3,-6.2)
  {AXI GPIO\\{\scriptsize 0xB100\_0000}};
\draw[ctl] (gpio.east) -- node[above, sub] {usr\_irq\_req} (3.9,-6.2);

\node[box, minimum width=4.0cm, minimum height=0.45cm] (ic) at (5.25,-7.45) {AXI Interconnect};
\draw[blue] (maxi.south) -- (maxi.south |- ic.north);
\draw[blue] (mbyp.south) -- (mbyp.south |- ic.north);
\node[box, minimum width=1.6cm, minimum height=0.6cm] (mig0) at (4.6,-8.55) {MIG\_0};
\draw[blue] (mig0.north |- ic.south) -- (mig0.north);
\node[box, fill=bdRam, text width=1.75cm, minimum height=0.7cm] (dimm0) at (4.6,-9.6)
  {PL DDR4\\{\scriptsize slots, shared buf.}};
\draw[link] (mig0) -- (dimm0);

\draw[verm] (mlite.east) -| (hpc0.south);
\node[sub, text=bdVerm] at (8.45,-6.65) {BAR0};
\draw[verm] (ic.east) -- (hpc1.south |- ic.east) -- (hpc1.south);

\node[xd, text width=2.0cm, minimum height=0.75cm] (x0) at (12.0,-8.55) {XDMA\_0\\{\scriptsize root port}};
\node[box, fill=black, text=white, minimum width=1.4cm, minimum height=0.6cm] (ssd0) at (12.0,-9.65) {SSD};
\draw[link] (x0) -- node[right, sub] {Gen3 x4} (ssd0);
\draw[blue] (x0.north) -- (12.0,-7.95) -- (5.9,-7.95) -- (5.9,-7.675);
\draw[verm] ($(x0.north)+(0.6,0)$) -- (12.6,-6.85) -- (13.6,-6.85) -- (hp.south);

\node[box, minimum width=2.6cm, minimum height=0.45cm] (icr) at (15.6,-7.0) {AXI Interconnect};
\draw[psm] (hpm0.south) -- (hpm0.south |- icr.north);
\node[box, dashed, text width=2.3cm, minimum height=0.75cm] (filt) at (15.6,-8.75)
  {4 $\times$ HLS filter\\{\scriptsize 333\,MHz, later rev.}};
\draw[psm] (icr.south) -- (filt.north);
\def\busy{-10.4}%
\draw[line width=0.6pt] (icr.west) -- (13.9,-7.0) -- (13.9,\busy) -- (1.3,\busy);
\draw[psm] (13.9,-8.55) -- (x0.east);
\draw[psm] (3.0,\busy) -- (3.0,-8.55) -- (mig0.west);
\draw[psm] (1.3,\busy) -- (gpio.south);
\foreach \x in {13.9,3.0} \fill (\x,\busy) circle (1.2pt);
\fill (13.9,-8.55) circle (1.2pt);
\node[sub, anchor=north] at (7.6,\busy) {M\_AXI\_HPM0\_FPD (PS master): control of XDMA\_0, AXI GPIO and HLS filters; PS access to PL DDR4};
\end{tikzpicture}
  \caption{\delilah detailed overview: (a)~software components, (b)~hardware components of the PL.
  \label{fig:overview}}
\end{figure*}

\subsection{Software Components}
\label{app:software}

Figure~\ref{fig:overview}(a) shows the software components of \delilah (white) on host and Zynq PS,
the components they rely on (coloured), and the application that uses them on the host (yellow).
Dark and light grey denote user space and kernel.

\paragraph{Host: API and driver.}
The application calls the \delilah API, which submits commands through io\_uring to the \delilah
driver\footnote{\url{https://github.com/delilah-csp/delilah-host}}.
The driver is a single kernel module that consists of two components, linked with the XDMA library: \emph{cdev},
which exposes a character device, and \emph{mod}, which registers the PCIe device, issues DMA transfers over
XDMA channels and writes command requests to BAR0.
The execution of a command on the driver proceeds as follows:
\begin{enumerate}
\item The user program sends an execute command to the \delilah driver through liburing.
\item cdev receives the submission queue entry, including relevant buffers and metadata, as well as the ranges of the
data slot to invalidate and flush (Section~\ref{sec:delilah:coherence}), and passes it to mod.
\item mod writes the command, including these parameters, to BAR0 (Appendix~\ref{app:memory}).
\item mod sets the command execute bit of BAR0.
\item Upon completion, the device writes the command response to BAR0 and raises an MSI-X interrupt.
The interrupt callback of mod, registered with XDMA, reads the command response and triggers a completion in liburing.
\end{enumerate}

\paragraph{Device: controller.}
The \delilah controller\footnote{\url{https://github.com/delilah-csp/delilah}} is a user-space program on the PS,
which runs on an embedded Linux image built with the PetaLinux configuration
of the \delilah project\footnote{\url{https://github.com/delilah-csp/delilah-pt}}.
It first configures the PCIe device and the eBPF run-time environment, then processes the commands sent by the host.
It maps BAR0, the program and data slots and the shared buffer into user space with
udmabuf\footnote{\url{https://github.com/delilah-csp/udmabuf}}
(magenta), writes the header of BAR0, and spawns a thread for each uBPF
virtual machine\footnote{\url{https://github.com/delilah-csp/ubpf}} (cyan), which allocates the virtual machine
state, registers the registered functions, and waits for commands from the host.
As the PS has 4 cores, we allocate four uBPF virtual machines, and pin the thread of virtual machine~$i$ to
core~$i$ (Figure~\ref{fig:memlayout}).

Each thread polls the command control register of its virtual machine. When the host requests the execution of
an eBPF program, the thread loads it from the program slot named in the command and executes it on the data slot
named in the command: a pointer to the data slot is passed as argument to the program, and the shared buffer is
accessible to all virtual machines.
The stack of each virtual machine has a fixed size, set when the controller is built. eBPF programs assume a
512-byte stack by default, which may be limiting for complex offloaded programs. We submitted a patch to LLVM, which now allows overriding
it~\cite{hedam_llvm_stack_2023}.

On completion of a command, the PCIe DMA/Bridge IP must receive a signal on a given pin.
The controller sends this signal through an AXI GPIO block, accessed through /dev/mem
(Appendix~\ref{app:hardware}), which XDMA\_2 turns into an MSI-X interrupt on the host;
the controller runs one IRQ thread per virtual machine.

\paragraph{Device: registered functions.}
Registered functions run in user space on the PS. They access the SSDs through the NVMe driver of the embedded Linux,
and the HLS filters (Appendix~\ref{app:hardware}) through UIO, using the registers exposed by the HLS-generated driver.
For the filters, five registered functions, one per predicate, are a facade for a single function that
translates the pointers of the eBPF program, which are relative to the data slot or the shared buffer, into physical
addresses; flushes the input range before starting the accelerator and invalidates the output range after completion,
restricting cache maintenance to the ranges it reads and writes; and drives the accelerator, polling for completion.

\subsection{Hardware Components}
\label{app:hardware}

\paragraph{Block design.}
Figure~\ref{fig:overview}(b) shows the components of the block design used in our
experiments\footnote{\url{https://github.com/delilah-csp/delilah-bd}}, built from Xilinx IP modules
and connected with each other and with the Zynq PS (grey) through AXI interconnects: paths to the PL DDR4
(thick solid blue), into the Zynq PS (dashed vermillion), and from the PS master port M\_AXI\_HPM0\_FPD (thin black).
The PCIe DMA/Bridge XDMA\_2 connects the device to the host over PCIe Gen3 x8 (8~GB/s); XDMA\_0, configured as a
PCIe root port, connects an M.2 SSD over PCIe Gen3 x4 (4~GB/s); the memory interface generator MIG\_0 controls a
DDR4 DIMM attached to the PL, which holds the program and data slots and the shared buffer.
The block design also instantiates a second SSD bridge and a second memory controller, which our experiments do not use
and which Figure~\ref{fig:overview}(b) omits.
Host DMA enters the PS through the coherent port S\_AXI\_HPC1\_FPD, host accesses to BAR0 through the AXI4-Lite
master M\_AXI\_LITE and S\_AXI\_HPC0\_FPD, and SSD DMA through the non-coherent port S\_AXI\_HP1\_FPD.
Through the master port M\_AXI\_HPM0\_FPD and an AXI interconnect, the device processors access the PL DDR4 through
MIG\_0, and control XDMA\_0, the AXI GPIO and the HLS filters.
The AXI GPIO drives the user interrupt lines of XDMA\_2, which the host receives as MSI-X interrupts.

\paragraph{HLS filters.}
The HLS filters\footnote{\url{https://github.com/delilah-csp/delilah-hw-filter}} (dashed in
Figure~\ref{fig:overview}(b)) belong to a later revision of the block design, which adds four instances of the filter,
clocked at 333~MHz, i.e., faster than the rest of the PL; their path to the PL DDR4 is not shown.
The filter supports five predicates (equal, not equal, less or equal, greater or equal, and between).
It reads its input and writes a position list through an AXI master port, and receives its parameters
(addresses, number of values, predicate and constants) through an AXI4-Lite control port.
It processes the input in vectors of 256 32-bit values staged in on-chip BRAM, so that it issues 1~KB memory
requests instead of one request per value, and compares the values of a vector in a pipelined loop.

\begin{figure}[htbp]
    \includegraphics[width=0.3\linewidth, height=5cm]{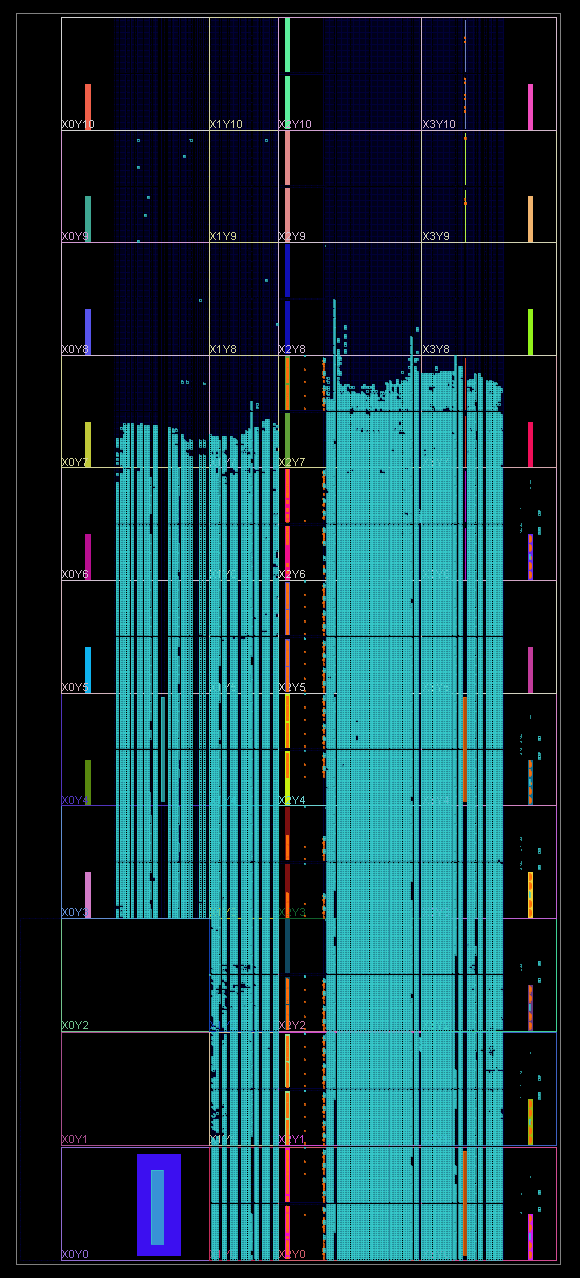}
    \includegraphics[width=0.3\linewidth, height=5cm]{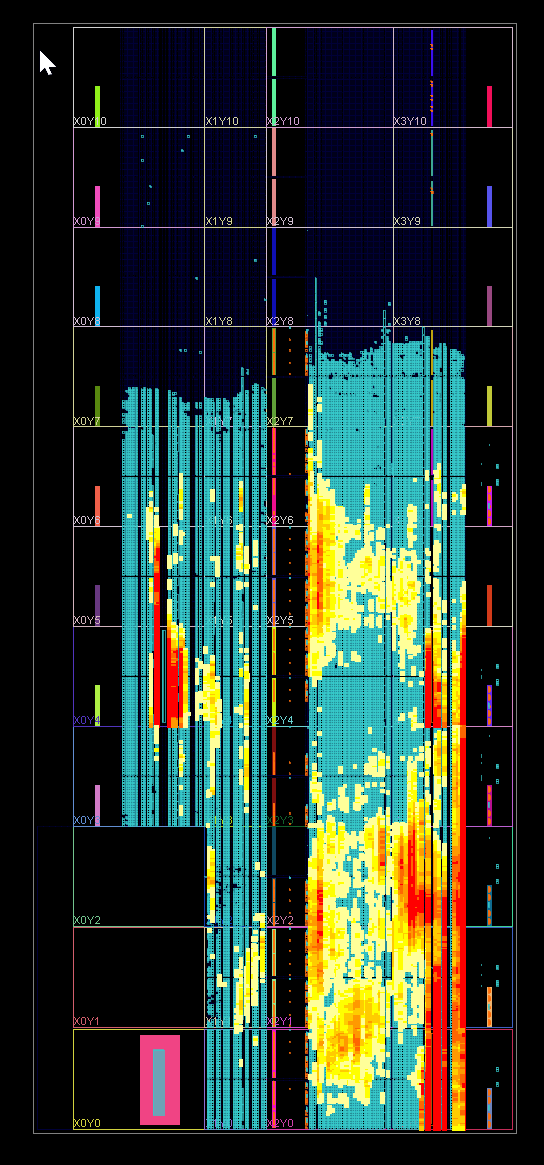}
    \includegraphics[width=0.3\linewidth, height=5cm]{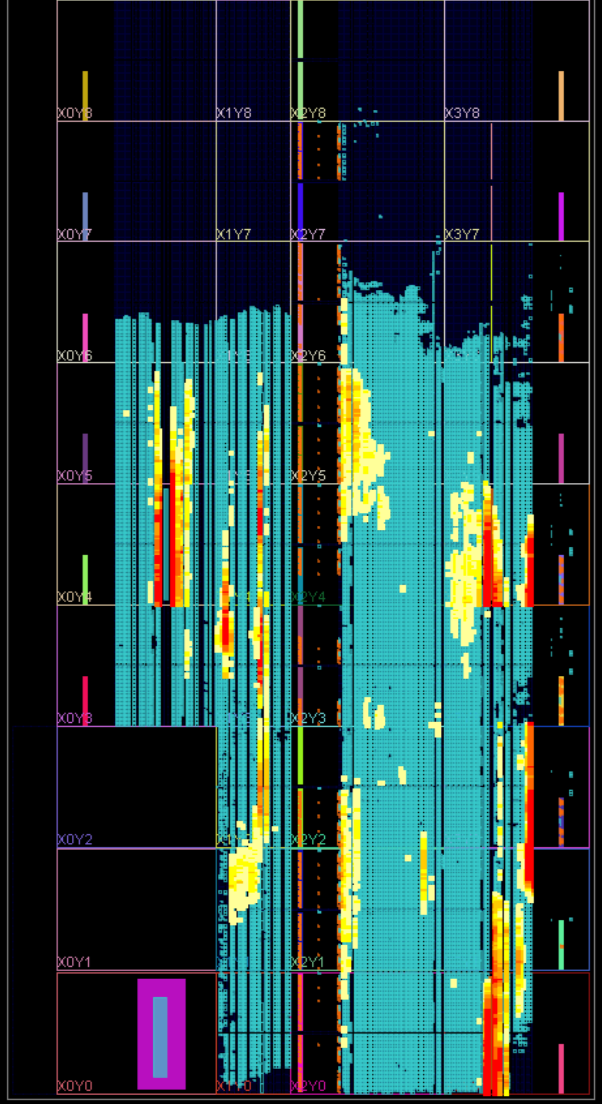}
    \caption{Top view of the Daisy FPGA (left), congested block design (middle) and block design without congestion (right).}
    \label{fig:fpga}
\end{figure}

\paragraph{Synthesis.}
The original version of our block design, with a 16-lane PCIe link to the host,
cannot be synthesized for the Daisy FPGA because of congestion issues.
The variant without PL DIMMs can be synthesized, but it takes 24 hours.
Figure~\ref{fig:fpga} shows the floor plan of the Daisy FPGA and an analysis
of the vertical congestion for two block designs.
When congestion exceeds 100\%, Vivado starts looking for alternate routes to complete routing.
Our original block design (middle) has maximum congestion of 325\% (red)
over a wide area, which causes routing to fail, because
no alternative route can be found that meets the timing constraints.
Reducing the number of PCIe lanes from 16 to 8 reduces congestion
significantly (right), which results in successful and much faster synthesis;
this is the block design of Figure~\ref{fig:overview}(b).
With 8 lanes, the host link still delivers 8~GB/s, more than the 3.2~GB/s that a DIMM
sustains when each byte is accessed four times in memory (Section~\ref{sec:delilah:design}).
A block design for writing data in bursts (e.g., writing objects in batch~\cite{do_better_2021}) might instead
favor host-to-memory bandwidth over SSD-to-memory bandwidth.

\subsection{Memory Layout}
\label{app:memory}

\begin{figure*}[tbp]
  \centering
  \definecolor{aUser}{HTML}{9E9E9E}
\definecolor{aKern}{HTML}{D3D3D3}
\definecolor{aApp}{HTML}{FFFF99}
\definecolor{aXdma}{HTML}{FFBFBF}
\definecolor{aUdma}{HTML}{FFBFFF}
\definecolor{aUbpf}{HTML}{BFFFFF}
\definecolor{aLnx}{HTML}{F2F2F2}
\definecolor{bdBlue}{HTML}{004488}
\definecolor{bdVerm}{HTML}{D55E00}
\definecolor{bdRam}{HTML}{FF8080}
\definecolor{bdPS}{HTML}{CCCCCC}
\definecolor{eZero}{HTML}{E69F00}
\definecolor{eOne}{HTML}{56B4E9}
\definecolor{eTwo}{HTML}{009E73}
\definecolor{eThree}{HTML}{CC79A7}
\begin{tikzpicture}[
  font=\footnotesize,
  box/.style={draw=black, line width=0.5pt, align=center, inner sep=2pt},
  app/.style={box, fill=aApp},
  dl/.style={box, fill=white},
  lnx/.style={box, fill=aLnx},
  xd/.style={box, fill=aXdma},
  ud/.style={box, fill=aUdma},
  vm/.style={box, fill=aUbpf},
  port/.style={box, fill=white, font=\scriptsize, inner sep=1pt},
  link/.style={draw=black, line width=0.6pt},
  zone/.style={font=\footnotesize\itshape},
  ptitle/.style={font=\footnotesize\bfseries, anchor=west, inner sep=0pt},
  title/.style={font=\footnotesize\bfseries},
  sub/.style={font=\scriptsize},
  blue/.style={draw=bdBlue, line width=1.6pt, -{Latex[length=2mm]}},
  verm/.style={draw=bdVerm, line width=1pt, dash pattern=on 3pt off 2pt, -{Latex[length=2mm]}},
  ctl/.style={draw=black, line width=0.5pt, -{Latex[length=1.6mm]}},
  row/.style={draw=black, line width=0.4pt},
]

\def\rh{0.3}%
\def\ybot{-16.45}%

\node[title, anchor=south, align=center] at (0.85,-11.65) {Host};
\draw[line width=0.6pt, fill=white] (0.1,\ybot) rectangle (1.6,-11.65);
\node at (0.85,-12.1) {$\vdots$};
\draw[line width=0.6pt, fill=aXdma] (0.1,-13.4) rectangle (1.6,-12.5);
\node[align=center] at (0.85,-12.95) {BAR0\\{\scriptsize 32\,MiB}};
\node at (0.85,-14.4) {$\vdots$};
\draw[line width=0.6pt, fill=aApp] (0.1,\ybot) rectangle (1.6,-15.35);
\node[align=center] at (0.85,-15.9) {DRAM\\{\scriptsize DMA}\\{\scriptsize buffers}};
\draw[densely dashed, line width=0.5pt] (1.6,-12.5) -- (2.3,-11.65) (1.6,-13.4) -- (2.3,\ybot);
\node[sub, rotate=90] at (1.95,-12.95) {MMIO};

\node[title, anchor=south, align=center] at (5.15,-11.65) {BAR0 region {\normalfont\scriptsize(PS DDR, via S\_AXI\_HPC0)}};
\newcommand{\barrow}[4]{%
  \draw[row, fill=#4] (2.3,\ybot+#1*\rh) rectangle (8.0,\ybot+#1*\rh+\rh);
  \node[font=\scriptsize\ttfamily, anchor=west, inner sep=1pt] at (2.35,\ybot+#1*\rh+0.5*\rh) {#2};
  \node[font=\scriptsize, anchor=west, inner sep=1pt] at (3.4,\ybot+#1*\rh+0.5*\rh) {#3};
}
\barrow{0}{0x0000}{ehver, ehbld[48]: version, build}{white}
\barrow{1}{0x0034}{eheng, ehpslot, ehdslot = 4, 16, 12}{white}
\barrow{2}{0x0038}{ehpsoff, ehpssze: program slots}{white}
\barrow{3}{0x0048}{ehdsoff, ehdssze: data slots}{white}
\barrow{4}{0x0058}{ehsssze: shared buffer size}{white}
\barrow{5}{}{\dots}{white}
\barrow{6}{0x1000}{commands[0]: request, response}{eZero!35}
\barrow{7}{0x1038}{commands[1]}{eOne!35}
\barrow{8}{0x1070}{commands[2]}{eTwo!35}
\barrow{9}{0x10A8}{commands[3]}{eThree!35}
\barrow{10}{}{\dots}{white}
\barrow{11}{0x2000}{cmdctrl[0]: exec, done}{eZero!35}
\barrow{12}{0x2007}{cmdctrl[1]}{eOne!35}
\barrow{13}{0x200E}{cmdctrl[2]}{eTwo!35}
\barrow{14}{0x2015}{cmdctrl[3]}{eThree!35}
\barrow{15}{}{\dots\ (up to 32\,MiB)}{white}
\draw[line width=0.6pt] (3.35,\ybot) -- (3.35,-11.65);
\draw[line width=0.8pt] (2.3,\ybot) rectangle (8.0,-11.65);

\node[title, anchor=south] at (9.95,-11.65) {Zynq PS cores};
\foreach \i/\c in {0/eZero,1/eOne,2/eTwo,3/eThree} {
  \node[box, fill=\c!35, minimum width=2.45cm, minimum height=1.05cm] (core\i) at (9.95,-15.9+1.2*\i) {};
  \node[font=\scriptsize, anchor=north] at (core\i.north) {Cortex-A53 core \i};
  \node[vm, font=\scriptsize, minimum width=1.75cm, inner sep=1pt] at ($(core\i.center)+(0,-0.15)$) {uBPF VM \i};
  \draw[draw=\c!80!black, line width=0.8pt] (8.0,\ybot+6*\rh+\i*\rh+0.5*\rh) -- (core\i.west);
}

\node[title, anchor=south, align=center] at (13.6,-11.65) {PL DDR4 {\normalfont\scriptsize(MIG\_0)}};
\draw[line width=0.6pt, fill=aUdma!60] (12.4,\ybot) rectangle (14.8,-15.25);
\foreach \k in {1,...,15} \draw[line width=0.3pt] (12.4,\ybot+\k*0.075) -- (12.6,\ybot+\k*0.075);
\node[align=center] at (13.7,-15.85) {program slots\\{\scriptsize 16 $\times$ 4\,MiB}};
\foreach \k/\t in {0/{data slot 0},1/{data slot 1},3/{data slot 11}} {
  \draw[line width=0.6pt, fill=aUdma!35] (12.4,-15.25+\k*0.6) rectangle (14.8,-14.65+\k*0.6);
  \node[font=\scriptsize] at (13.6,-14.95+\k*0.6) {\t};
}
\draw[line width=0.6pt, fill=aUdma!35] (12.4,-14.05) rectangle (14.8,-13.45);
\node at (13.6,-13.7) {$\vdots$};
\draw[line width=0.6pt, fill=bdRam!35] (12.4,-12.85) rectangle (14.8,-11.65);
\node[align=center] at (13.6,-12.25) {shared buffer\\{\scriptsize 1\,GiB, device only}};
\draw[line width=0.8pt] (12.4,\ybot) rectangle (14.8,-11.65);
\foreach \y/\a in {\ybot/{0x10\_0000\_0000}, -15.25/{0x10\_0400\_0000}, -12.85/{0x13\_0400\_0000}, -11.65/{0x13\_4400\_0000}}
  \node[font=\scriptsize\ttfamily, anchor=west, inner sep=1pt] at (14.85,\y) {\a};
\node[sub, anchor=west, align=left] at (14.9,-13.95) {data slots\\12 $\times$ 1\,GiB};

\draw[line width=0.6pt] (11.4,-16.3) -- (11.4,-11.85);
\foreach \i in {0,1,2,3} \draw[line width=0.6pt] (core\i.east) -- (11.4,-15.9+1.2*\i);
\draw[ctl] (11.4,-15.85) -- node[above, sub] {prog.\ $p$} (12.4,-15.85);
\draw[ctl] (11.4,-14.35) -- node[above, sub] {data $d$} (12.4,-14.35);
\draw[ctl] (11.4,-12.25) -- node[above, sub] {all} (12.4,-12.25);

\draw[draw=black, line width=0.9pt, {Latex[length=2mm]}-{Latex[length=2mm]}]
  (0.85,\ybot) -- (0.85,-16.95) -- node[below, sub] {XDMA H2C/C2H DMA to ehpsoff\,+\,$p\cdot$ehpssze (program slot $p$) and ehdsoff\,+\,$d\cdot$ehdssze (data slot $d$)} (13.6,-16.95) -- (13.6,\ybot);
\node[sub, anchor=east, font=\scriptsize\itshape] at (17.0,-17.35) {not to scale};
\end{tikzpicture}
  \caption{\delilah memory layout: BAR0, program and data slots, shared buffer, and placement of the uBPF virtual
  machines on the PS cores. \label{fig:memlayout}}
\end{figure*}
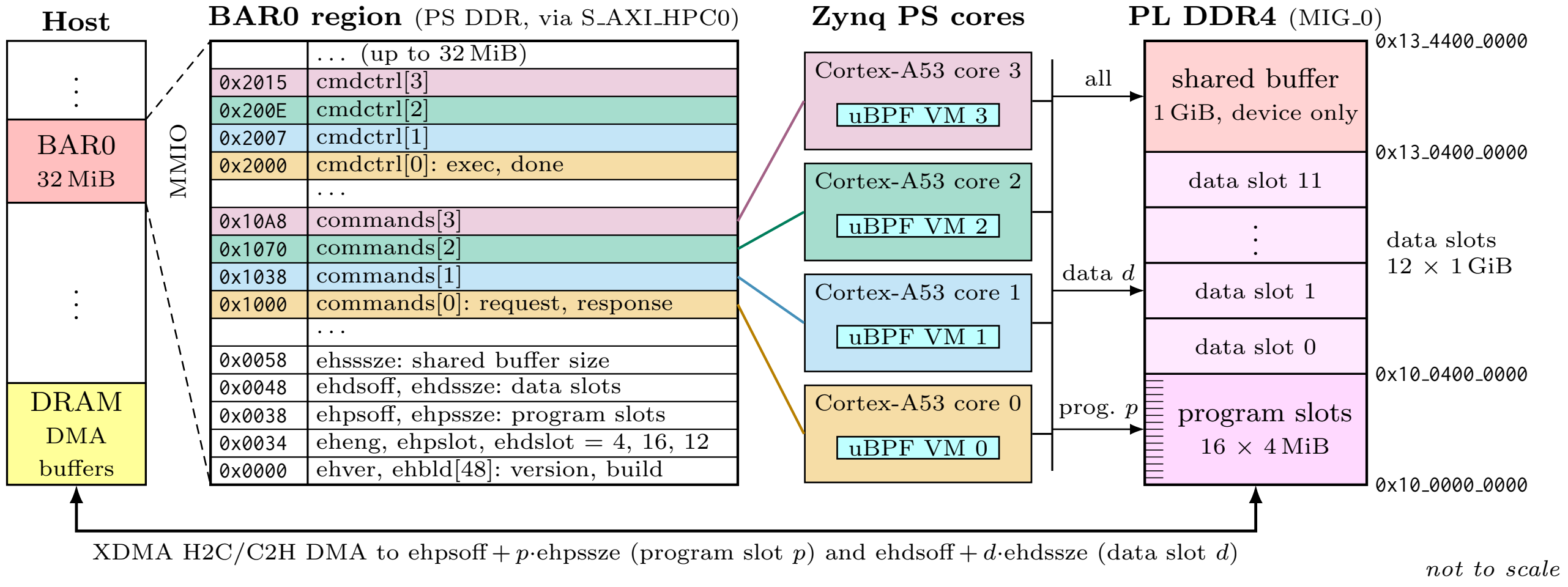

Figure~\ref{fig:memlayout} shows BAR0, the program and data slots and the shared buffer in device memory,
and the placement of the uBPF virtual machines on the PS cores.

\paragraph{BAR0.}
BAR0 is the PCIe base address register that contains the PCIe device id and defines a non-prefetchable 32~MiB aperture,
memory-mapped in the host address space. It resides in the PS DDR, and the controller maps it into user space.
It holds:
\begin{itemize}
    \item a header, written by the controller when it starts, with the number of uBPF run-time environments, i.e.,
    the number of uBPF virtual machines spawned by the controller, the number, base address and size of the program
    slots and of the data slots, and the size of the shared buffer;
    \item for each uBPF run-time environment:
    \begin{itemize}
        \item a command (commands[$i$]), written by the host, composed of an opcode, an id and command-specific
        configuration bits (for the execution command, the program slot, the data slot, the program length, and the
        ranges of the data slot to invalidate and flush);
        \item a response, written by the device, containing a status code when the command is executed;
        \item a command control register (cmdctrl[$i$], with exec and done bits), used by the host to signal the device
        that a command is ready to be processed and by the device to signal that the command has been executed.
    \end{itemize}
\end{itemize}

\paragraph{Slots and shared buffer.}
The program and data slots and the shared buffer reside in the PL DDR4 behind MIG\_0.
The host computes the device address of a slot from the header, e.g., the base address of the data slots plus
$d$ times their size for data slot~$d$, and transfers programs, arguments and results by DMA to and from that address.
The shared buffer is only accessible from the device.
The worker thread of uBPF virtual machine~$i$ is pinned to core~$i$, polls cmdctrl[$i$], and executes the program
slot~$p$ on the data slot~$d$ named in commands[$i$].
Because the host and the SSD access the slots by DMA without snooping the caches of the device processors,
the controller keeps them coherent in software (Section~\ref{sec:delilah:coherence}).

\end{document}